%% file: paper.tex
\documentclass[preprint,aip,superscriptaddress,preprintnumbers,amsmath,amssymb]{revtex4-1}
\usepackage{graphicx}
\usepackage[version=2]{mhchem}
\usepackage{array}
\usepackage{siunitx}
\usepackage{booktabs}
\usepackage{gensymb}
\usepackage{multirow}
\usepackage{dcolumn} 
\usepackage{setspace}
\usepackage{array}
\usepackage{psfrag}
\usepackage{textcomp}
\usepackage{color}
\usepackage{makecell}
\usepackage{listings}
\usepackage{subfig}
\usepackage{floatrow}
\usepackage{hyperref}
\hypersetup{colorlinks=true,
	linkcolor=blue,
	filecolor=black,      
	urlcolor=black,
	citecolor=blue,
	anchorcolor=black}

\input{macros}

\graphicspath{{figures/}}

\begin{document}

\title{DFTpy: An efficient and object-oriented platform for orbital-free DFT simulations}

\author{Xuecheng Shao}
\email{xuecheng.shao@rutgers.edu}
\affiliation{Department of Chemistry, 73 Warren St., Rutgers University, Newark, NJ 07102, USA}

\author{Kaili Jiang}
\email{kaili.jiang@rutgers.edu}
\affiliation{Department of Chemistry, 73 Warren St., Rutgers University, Newark, NJ 07102, USA}

\author{Wenhui Mi}
\email{wenhui.mi@rutgers.edu}
\affiliation{Department of Chemistry, 73 Warren St., Rutgers University, Newark, NJ 07102, USA}

\author{Alessandro Genova} 
\altaffiliation{Current address: Kitware Inc., 1712 U.S. 9 Suite 300, Clifton Park, NY 12065, USA}
\affiliation{Department of Chemistry, 73 Warren St., Rutgers University, Newark, NJ 07102, USA}

\author{Michele Pavanello}
\email{m.pavanello@rutgers.edu}
\affiliation{Department of Chemistry, 73 Warren St., Rutgers University, Newark, NJ 07102, USA}
\affiliation{Department of Physics, 101 Warren St., Rutgers University, Newark, NJ 07102, USA}

\begin{abstract}
In silico materials design is hampered by the computational complexity of Kohn-Sham DFT, which scales cubically with the system size. Owing to the development of new-generation kinetic energy density functionals (KEDFs), orbital-free DFT (OFDFT, a linear-scaling method) can now be successfully applied to a large class of semiconductors and such finite systems as quantum dots and metal clusters. In this work, we present DFTpy, an open source software implementing OFDFT written entirely in Python 3 and outsourcing the computationally expensive operations to third-party modules, such as NumPy and SciPy. When fast simulations are in order, DFTpy exploits the fast Fourier transforms (FFTs) from PyFFTW. New-generation, nonlocal and density-dependent-kernel KEDFs are made computationally efficient by employing linear splines and other methods for fast kernel builds. We showcase DFTpy by solving for the electronic structure of a million-atom system of aluminum metal which was computed on a single CPU. The Python 3 implementation is object-oriented, opening the door to easy implementation of new features. As an example, we present a time-dependent OFDFT implementation (hydrodynamic DFT) which we use to compute the spectra of small metal cluster recovering qualitatively the time-dependent Kohn-Sham DFT result. The Python code base allows for easy implementation of APIs. We showcase the combination of DFTpy and ASE for molecular dynamics simulations (NVT) of liquid metals. DFTpy is released under the MIT license.
\end{abstract}

\date{\today}
\maketitle

\section{Introduction}
\input{intro.tex}

\section{Classes and software workflow}

\input{classes.tex}
\section{Details enabling computational efficiency}
\input{details.tex}

\label{details}
\section{Timings and assessment of efficiency}
\input{applications.tex}

\label{applications}
\section{Ease of implementation of new methods}
\input{ease.tex}

\section{Conclusions And Future Directions}
The Python revolution in computational electronic structure theory began almost two decades ago. It initially involved the emergence of wrappers for traditional software \cite{Hjorth_Larsen_2017,Ong_2013,Jacob_2011}. Initial attempts to output full-fledged quantum chemistry implementations came as early as 2004 \cite{pyquante}. A defining moment was the 2015 release of PySCF \cite{PYSCF} which featured an essentially complete quantum chemistry code (including advanced post-HF methods) with a software that leveraged C routines for Gaussian integrals and Python for essentially anything else.

With DFTpy, we merely follow this revolution, by developing a Python implementation for OFDFT simulations. The object-oriented nature of DFTpy provides an almost barrierless entry to advanced coding. We give an example of this by showcasing a new time-dependent OFDFT implementation and associated applications to the optical spectra of Mg clusters. In addition to the clear advantages compared to other, more traditional OFDFT codes based on low-level programming languages, DFTpy implements new-generation nonlocal KEDFs with density dependent kernels in a fairly efficient way. An analysis of timings shows that the cost associated with the new nonlocal functionals is less than 20 times that of semilocal functionals when isolated systems are approached (such as surfaces or clusters) and less than 7 times when bulk systems are considered. This is an important advance, making such functionals feasible for large scale simulations.

DFTpy classes and structure are general and could support a KS-DFT implementation and APIs to other Python codebases, such as PySCF, GPAW and PSI4. In doing so, in the near future we will implement a set of classes that will handle embedding schemes (from many-body expansions to density and quantum embedding). In this way, we will be able to seamlessly combine portions of a mesoscopic system computed at the OFDFT level and others at the KS-DFT level pushing the boundaries of time and length scales that can be approached by {\it ab initio} methods.

\begin{acknowledgments} 
This material is based upon work supported by the National Science Foundation under Grant No.\ CHE-1553993. The TD-OFDFT development is supported by the U.S.\ Department of Energy, Office of Basic Energy Sciences, under Award Number DE-SC0018343.

The authors acknowledge the Office of Advanced Research Computing (OARC) at Rutgers, The State University of New Jersey for providing access to the Amarel cluster and associated research computing resources that have contributed to the results reported here. URL: http://oarc.rutgers.edu 
\end{acknowledgments}
\bibliography{DFTpy}
\end{document}

%% file: macros.tex
\usepackage{amsmath}
\DeclareMathOperator*{\argmin}{argmin}

\newcommand{\eqn}[1]{\mbox{Eq.\hspace{1pt}(\ref{#1})}}
\newcommand{\eqs}[2]{\mbox{Eq.\hspace{1pt}(\ref{#1}--\ref{#2})}}

\newcommand{\eqtn}[2]{\begin{equation} \label{#1} #2 \end{equation}}

\def\brp{{\mathbf{r}^{\prime}}}

\def\bj{{\mathbf{j}}}

\def\br{{\mathbf{r}}}

\def\d{{\mathrm{d}}}

\def\vv{{von Weizs\"acker}}

\def\diff{{\rm d}}

\def\sss{\scriptscriptstyle\rm}

\def\s{_{\sss S}}

\def\Pauli{^{\rm Pauli}}

\def\VW{^{\rm vW}}

\usepackage[utf8]{inputenc}
 
\usepackage{listings}
\usepackage{xcolor}
 
\definecolor{codegreen}{rgb}{0,0.6,0}
\definecolor{codegray}{rgb}{0.5,0.5,0.5}
\definecolor{codepurple}{rgb}{0.58,0,0.82}
\definecolor{backcolour}{rgb}{0.95,0.95,0.92}
\lstdefinestyle{mystyle}{
    backgroundcolor=\color{backcolour},   
    commentstyle=\color{codegreen},
    keywordstyle=\color{magenta},
    numberstyle=\tiny\color{codegray},
    stringstyle=\color{codepurple},
    basicstyle=\ttfamily\footnotesize,
    breakatwhitespace=false,         
    breaklines=true,                 
    captionpos=b,                    
    keepspaces=true,                 
    numbers=left,                    
    numbersep=5pt,                  
    showspaces=false,                
    showstringspaces=false,
    showtabs=false,                  
    tabsize=2
}
 

%% file: intro.tex
\subsection{Theoretical background}
Orbital-free Density Functional Theory (OFDFT) is an emerging technique for modeling materials (bulk and nanoparticles) with an accuracy nearing the one of Kohn-Sham DFT (KSDFT) and with an algorithm that is almost linear scaling, $O(Nlog(N))$, both in terms of work and memory \cite{witt2018orbital,Chen2015228,shao2018large}. The most efficient OFDFT software \cite{Chen2015228,gavini2007non,Chen_2016,shao2018large} can approach million-atom system sizes while still accounting for the totality of the valence electrons. The central ingredient to OFDFT is the employment of pure Kinetic Energy Density Functionals (KEDFs). Commonly adopted KEDF approximants are not accurate enough to describe strongly directional chemical bonds -- a category which unfortunately includes most molecules \cite{xia2012,xia2012b}. However, new-generation nonlocal KEDFs allow OFDFT to model quantum dots and semiconductors \cite{mi2019LMGP,Xu_2019}. Hence, OFDFT is to be considered an emerging technique for computational materials science, chemistry and physics.

In OFDFT, the electronic structure is found by direct minimization of the DFT Lagrangian,
\eqtn{lag}{\mathcal{L}[\rho]=E[\rho]-\mu\left(\int\rho(\br) \mathrm{d} \br - N_{e} \right),}
where $E[\rho]$ is the electronic energy density functional, and $N_e$ the number of valence electrons, taking the form,
\eqtn{efunc}{
E[\rho] = T_s[\rho] + E_{H}[\rho] + E_{xc}[\rho] + \int v_\text{ext}(\br) \rho(\br) \d\br,
}
where, $E_H$ is the Hartree energy, $E_{xc}$ the exchange-correlation (xc) energy, $T_s$ the noninteracting kinetic energy and $v_\text{ext}(\br)$ is the external potential (in OFDFT, typically given by local pseudopotentials). 

Minimization of the Lagrangian with respect to the electron density function, $\rho(\br)$, yields the density of the ground state. In other words,
\eqtn{lag1}{\rho(\br)=\argmin_{\rho}\left\{\mathcal{L}[\rho]\right\}.}
In practice, $\rho(\br)$ can be obtained by solving the Euler-Lagrange equation,
\eqtn{euler}{
\frac{\delta E[\rho]}{\delta \rho(\br)}-\mu=0,
}
which is expanded as follows
\eqtn{euler2a}{
\frac{\delta T_s[\rho]}{\delta \rho(\br)} + v_s(\br)-\mu=0,
}
where we grouped $v_s(\br) = \frac{\delta E_H[\rho]}{\delta \rho(\br)} + \frac{\delta E_{xc}[\rho]}{\delta \rho(\br)} +v_\text{ext}(\br)$.

In conventional KSDFT, the KEDF potential, $\frac{\delta T_s[\rho]}{\delta \rho(\br)}$, is not evaluated and instead the kinetic energy is assumed to be only a functional of the KS orbitals which in turn are functionals of the electron density. In OFDFT, the KEDF potential is available by direct evaluation of the functional derivative of an approximate KEDF. Thus, the Euler equation \eqn{euler2a} can be tackled directly.

\subsection{OFDFT software background}
In this work, we present DFTpy, a flexible and object-oriented implementation of OFDFT. The software builds all the needed energy and potential terms so that the minimization of the energy functional can be carried out. The optimization itself can be done by several commonly adopted nonlinear, multi variable optimizers (such as quasi-Newton methods).

DFTpy situates itself in a fairly uncultivated field, as unlike KSDFT, OFDFT software are few \cite{Chen2015228,gavini2007non,Lehtom_ki_2014,Karasiev_2014,ATLAS}. As most projects, DFTpy started out as a toy project collecting Python 3 classes defining {\tt NumPy.Array} subclasses and associated methods for handling functions on regular grids. Functionalities included interpolations and conversion between file types. This was released under the moniker PBCpy \cite{pbcpy}. The next step for DFTpy came in 2018 when classes related to the basic energy terms in materials were developed. Hartree energy based on NumPy's FFTs, exchange-correlation and KEDF functionals based on pyLibXC \cite{Lehtola_2018}. Efforts to formalize the previous implementation culminated in recent months with a strong focus on efficiency of the codebase for its application to million atom systems. 

The current state-of-the-art in OFDFT software is PROFESS \cite{Chen2015228}, GPAW \cite{Lehtom_ki_2014}, ATLAS \cite{ATLAS} and DFT-FE \cite{gavini2007non}. GPAW, DFT-FE and ATLAS are real-space codes implementing either finite-element or finite-difference methods. Similarly to PROFESS, DFTpy relies on Fourier space not only for the treatment of Coulomb interactions but also for the computation of gradient and Lapacian operations (needed for instance for the \vv\ term). 


The distinguishing new features of DFTpy lie in its object-oriented core design composed of several important abstractions: {\tt Grid}, {\tt Field} (i.e., functions on grids), {\tt FunctionalClass} (i.e., an abstraction encoding an energy functional). These enable fast implementations of new functionalities. As an example, in this work we showcase a new time-dependent OFDFT \cite{Banerjee_2000,Tokatly_1999,Banerjee_2008,Zaremba_1994} implementation for the computation of optical spectra within an OFDFT framework, and an API combining DFTpy with ASE \cite{Hjorth_Larsen_2017} for the realization of molecular dynamics simulations.

More specifically, DFTpy distills efficient methods for the computation of structure factors {\it via} the smooth particle-mesh Ewald method \cite{Darden_1993,Essmann}, and an in-house, line-search-based electron density optimization algorithm which has the ability to dynamically adjust the effective grid cutoff during the optimization. To our knowledge, DFTpy contains the most efficient implementation to date of new-generation nonlocal KEDFs. These functionals are known to give a major boost to the performance of semilocal and nonlocal KEDFs but are associated with an unsustainable increase in the computational cost. DFTpy solves the problem by implementing an evaluation of the KEDF functional derivative (potential) that exploits linear splines, bringing down the computational cost to less than 20 times the one of a GGA KEDF.

The paper is organized as follows. We first describe the most important classes defining the DFTpy codebase. We proceed to describe the core aspects responsible for the efficient implementation. Lastly, we provide the reader with two examples: first showcasing DFTpy timings and linear scalability with system size and then DFTpy's ease of implementation of new methods by presenting a time-dependent OFDFT implementation that we apply to the computation of optical spectra of small metallic clusters.

%% file: classes.tex
\subsection{DFTpy classes}
DFTpy bases itself on PBCpy, a collection of classes for handling functions and fields of arbitrary rank in periodic boundary conditions \cite{pbcpy}. PBCpy's main classes are {\tt Grid} and {\tt Field} which are both {\tt NumPy.Array} subclasses.

The {\tt Grid} class (comprising of {\tt BaseGrid}, {\tt DirectGrid} and {\tt ReciprocalGrid}) is aware of all the attributes needed to define a grid, such as the lattice vectors, the number of space discretization points in each direction. Subclasses of {\tt Grid} include {\tt RealSpaceGrid} and {\tt ReciprocalSpaceGrid} which are self-explanatory. 

The {\tt Field} class (comprising of {\tt BaseField}, {\tt DirectField} and {\tt ReciprocalField}) encodes a function defined on a {\tt Grid}. There are several methods bound to this class. For example, if {\tt Field} is defined on a {\tt Direct}/{\tt ReciprocalGrid} it contains {\tt .fft}/{\tt .ifft}, the forward/inverse Fourier Transform. Additional bound methods include spline interpolations, integrals, and the appropriately extended definitions of the common algebraic operations ({\tt =, +, *, /}). Fields can be of arbitrary rank. For instance, the electron density is a rank one field, while the density gradient is a rank three field whether they are represented in real or reciprocal space.

DFTpy features classes, such as {\tt FunctionalClass} for the evaluation of the various terms in the energy: the kinetic energy density functional, {\tt KEDF}, the exchange-correlation functional, {\tt XC}, the electron-ion local pseudopotential, {\tt IONS}, and the Hartree functional, {\tt HARTREE}. Such a software structure is compatible with virtually all types of electronic structure methods, and not only OFDFT. Thus, we expect in future releases of DFTpy to also include KSDFT as well as APIs at the level of the energy functional for external KSDFT codes and particularly those offering efficient Python interfaces\cite{PYSCF,Smith_2018,Lehtom_ki_2014}.

\begin{figure}[htp]
	\begin{tabular}{cc}
		\sidesubfloat[]{\includegraphics[width=0.45\textwidth]{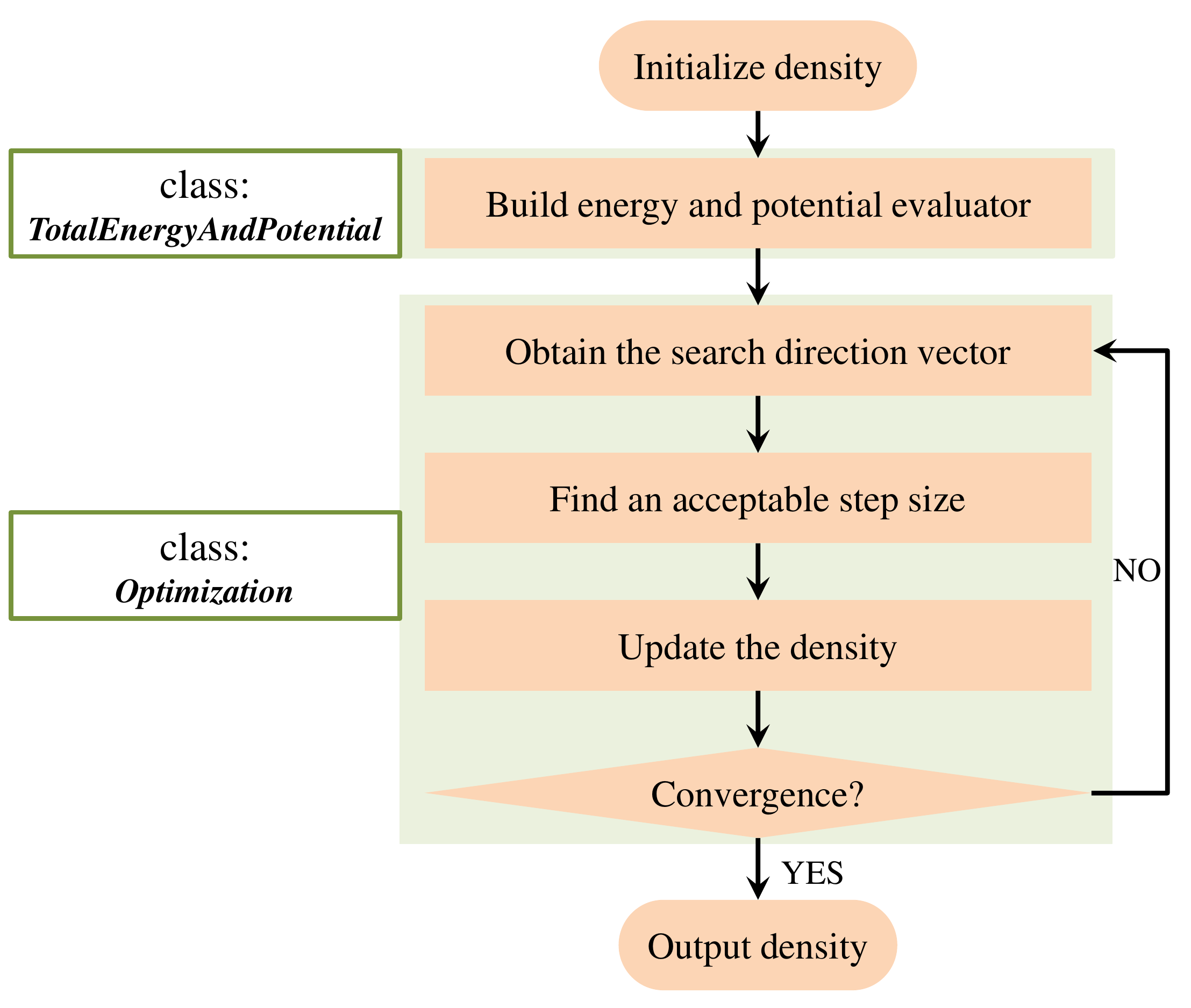}\label{fig:flow_a}} \hspace{1em}
		\sidesubfloat[]{\includegraphics[width=0.45\textwidth]{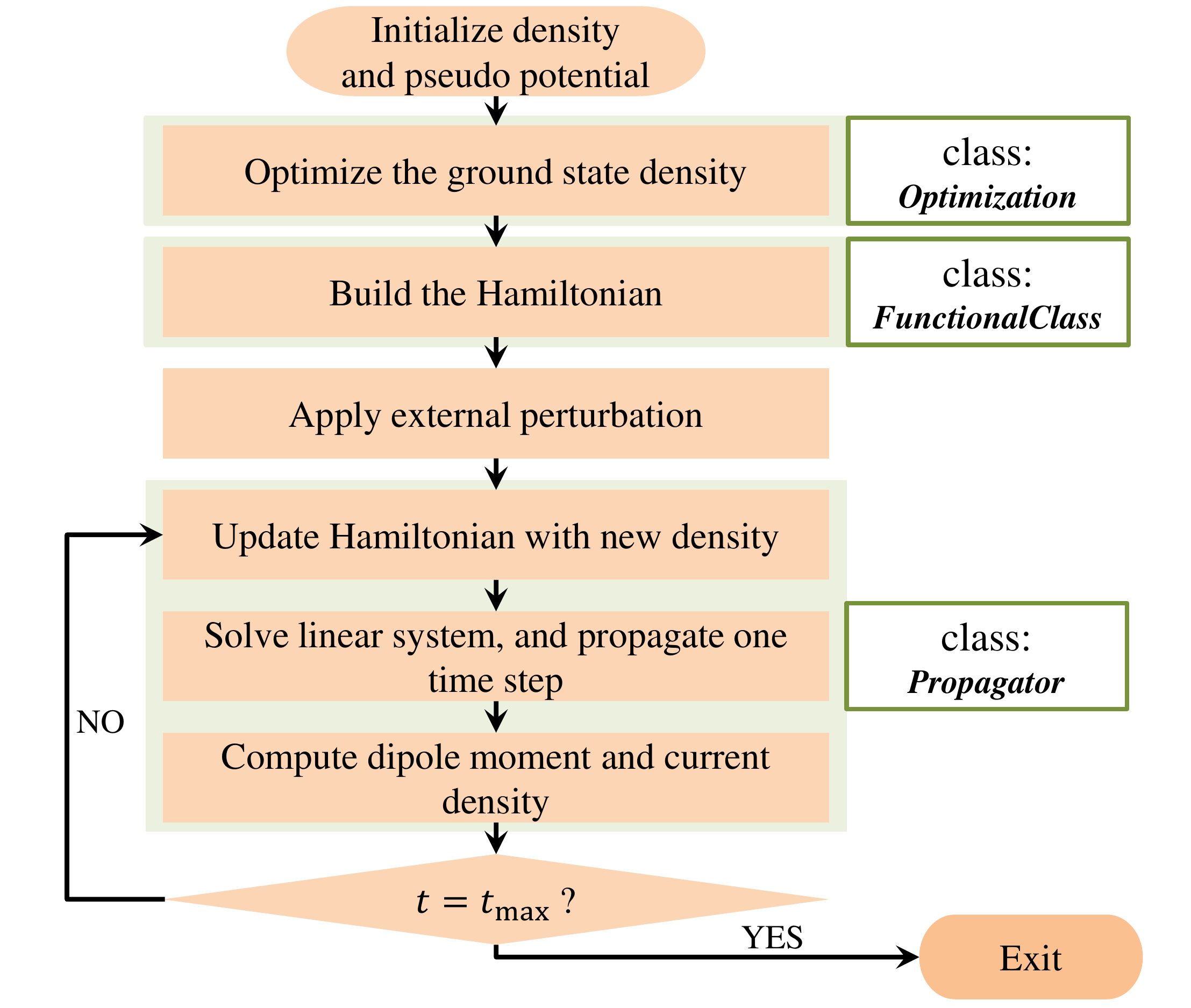}\label{fig:flow_b}}
	\end{tabular}
\caption{\label{fig:flow}Flowcharts for (a) a density optimization job, and (b) a TD-OFDFT job (see text for details). On the side of the flowcharts, we indicate in green boxes the names of the Python classes involved.}
\end{figure}

\subsection{Other Classes and APIs}

DFTpy contains classes for handling the optimization of the electron density and for handling the user interface. The optimization class is a standard optimizer which will probably be spun off as its own module in later releases. The user interface consists of a Python dictionary collecting the parameters for the calculation and an API to ASE's input/output geometry handler \cite{Hjorth_Larsen_2017}. With ASE, DFTpy can read and write virtually any file format.

DFTpy has been conceived to ease developments of new methods and to leverage the many modules already available. Too often junior scientists spend time reinventing common software simply because their platform is not flexible enough to interface easily with other modules. 
We showcase this with a simple example, using the capability of ASE to run molecular dynamics with DFTpy as the external engine. We developed a {\tt DFTpyCalculator} class which is in the form of an ASE {\tt Calculator} class, set in the {\tt ASE.Atoms} class. In Section \ref{details}, we present a simple example of MD simulation carried out with DFTpy+ASE. 

Two workflow examples are given in Figure \ref{fig:flow}. In inset \subref{fig:flow_a} of the figure, we show a flowchart describing the main steps carried out by a density optimization job. Only 3 classes are involved: {\tt FunctionalClass}, {\tt EnergyAndPotentialEvaluator} and {\tt Optimization}. Figure \ref{fig:flow}\subref{fig:flow_b} shows the flowchart of TF-OFDFT job. For a TD-OFDFT calculation ({\it vide infra}), an additional class is required. Namely, {\tt Propagator} needed for handling the TDDFT propagation step. Examples and Jupyter Notebooks related to the density optimization class and the DFTpy+ASE API are available at DFTpy's manual \cite{dftpyweb} and git repository \cite{dftpygit}.

%% file: details.tex
\subsection{New-generation nonlocal KEDFs}
New-generation nonlocal KEDFs began with the breakthrough development of the Huang-Carter functional (HC) in 2010 \cite{PhysRevB.81.045206}. For the first time, this functional could reliably approach semiconductors and inhomogeneous systems with a robust algorithm. Unfortunately, HC was deemed too computationally expensive to become a workhorse for realistically sized model systems. This prompted a number of additional development by several groups \cite{Constantin_2018,luo2018simple,constantin2017,xu2020nonlocal}, including our recent work \cite{mi2018nonlocal,mi2019LMGP}. In this section, we will focus on the functionals developed by our group, and specifically LMGP and LWT family of functionals. However, the techniques and conclusions drawn here are general and encompass other new-generation functionals, such as HC \cite{PhysRevB.81.045206} and LDAK \cite{xu2020nonlocal}.

Nonlocal KEDFs share the form:  
\begin{equation}
\label{nl2}
T_{NL}[\rho]=\int{\rho^{\alpha}(\br)\omega_{NL}[\rho]({\br, \brp})\rho^{\beta}(\brp)d{\br}d{\brp}},
\end{equation}
where $\alpha$ and $\beta$ are suitable parameters, and $\omega_{NL}[\rho]({\br, \brp})$ is a kernel usually assumed to be a function of only $|\br-\brp|$ and as such is represented in reciprocal space by a one-dimensional function, $\omega_{NL}(q)$. When the Wang-Teter functional is used \cite{PhysRevB.45.13196},
\begin{align}
\label{WT}
\omega_{NL}(q)=\omega_{\rm WT}(q)={\rm C_{WT}}\,G_{NL}(\eta(q))
\end{align}
where $\eta(q)=\frac{q}{2k_F}$ with $k_F=(3\pi^2\rho)^{\frac{1}{3}}$ is the Fermi wavevector, and $ \rm C_{WT}=\frac{6}{25}{(3\pi^{2})^{2/3}} $.
The WT functional can be improved to satisfy functional integration relations \cite{mi2018nonlocal} by the addition of one correction term giving rise to the MGP family of functionals. Namely,
\begin{align}
\label{MGP}
\omega_{x,y}(q)=\omega_{\rm WT}(q)-x{\rm C_{WT}}\int_{0}^{1}dt ~t^{y}\frac{d G_{NL}(\eta(q,t))}{dt}.
\end{align}
%
where
\begin{equation}
G_{NL}(\eta)=\left( \frac{1}{2} + \frac{1-\eta^2}{4\eta} \ln \left| \frac{1+\eta}{1-\eta} \right| \right)^{-1} - 3\eta^{2} -1.
\end{equation}
MGP is given by $(x,y)=\left(1, 5/6\right)$, MGPA by $(x,y)=\left(1/2, 5/6\right)$ and MGPG by $(x,y)=\left(1, 5/3\right)$. The only difference between MGP/A/G is the way a kernel is symmetrized. We refer the interested reader to the supplementary information of Ref.\ \citenum{mi2018nonlocal}. 

In Ref.\ \citenum{mi2019LMGP} we developed a technique to generalize WT as well as  MGP/A/G functionals to approach localized, finite systems by invoking spline techniques to obtain kernels no longer dependent only on the average electron density but instead dependent locally on the full electron density function. The resulting functionals are dubbed LWT, LMGP/A/G depending on the kernels mentioned in \eqs{WT}{MGP}. 

The implementation of these new functionals in DFTpy requires the following four steps (s1-s4):

\begin{itemize}
\item[s1] Determine the maximum/minimum value of $k_F$ and generate a set of $k_F$ values in the $\left[k_F^\text{max}, k_F^\text{min}\right]$ interval by an arithmetic or geometric  progression. This is an $O(N)$ operation with a very small prefactor. 
\item[s2] Evaluate the kernel for each of the $k_F$s using splines either in real or reciprocal space. At the beginning of computation, the kernel is calculated at some discrete points of $\eta$. This calculation is done only once. The kernel evaluation is an $O(N)$ operation with a potentially large prefactor depending on the type of spline used.
\item[s3] Compute the KEDF potential from \eqn{nl2} with the  different kernels. This is an $O(N\log N)$ operation due to the FFTs needed to evaluate the convolution in \eqn{nl2}.
\item[s4] Interpolate the KEDF potential over the values of $k_F$ onto $k_F[\rho(\br)]$ to obtain the final nonlocal KEDF potential and energy. This is an $O(N)$ operation with a potentially large prefactor depending on the type of spline used.
\end{itemize}

The timings associated with $T_{NL}$ compared to $T_{\rm TF}+T_{\rm vW}$ are summarized in Table \ref{tlmgp} for Aluminum clusters of different sizes ranging from 13 to 12,195 atoms. The structures are generated with {\tt ASE} adopting a 15 \AA{} vacuum layer in each direction to ensure the interactions between atoms and their periodic images are negligible.  Inspecting Table \ref{tlmgp}, we note that the cost of the additional nonlocal functionals is less than 20 times the one of the semilocal functionals. Thus, it is reasonable to conclude that DFTpy's implementation of new-generation nonlocal KEDFs opens the door to predictive, {\it ab initio} simulations of mesoscale systems ($>10$ nm).

\begin{table}
	\caption{\label{tlmgp}Timings for the evaluation of new-generation KEDFs with DFTpy for Al clusters of varying sizes. 40 $k_{F}$ values between $k_F^\text{min}$ and $k_F^\text{max}$ generated with an arithmetic progression is used in all systems. The kernel is interpolated using the nearest-neighbor method and the linear spline is employed to interpolate the nonlocal KEDF potential over the $k_F$ values.}
	\begin{tabular}{c|cccc}
		\hline
		        \# atoms         &  13   &  171   &  1099  &  12195  \\ \hline
		       $ T_{NL} $        & 87.14 & 227.22 & 759.64 & 8010.22 \\
		$ T_{\rm TF}+T_{\rm vW}$ & 6.15  & 16.54  & 50.83  & 463.36  \\ \hline
	\end{tabular}
\end{table}

We should make the following remarks: (1) The results presented in Table \ref{tlmgp} are a reference only to isolated systems. For bulk systems, a much smaller $k_F$ grid is needed and the cost is therefore much reduced. Testing shows that the cost becomes less than half of the one in the table for similarly sized bulk systems. (2) The arithmetic progression used to generate the $\eta$ grid can be improved and optimized. For example, we found that using geometric progressions can reduce the number of needed $\eta$ points and thus further reduce the cost compared to Table \ref{tlmgp}.

\subsection{Density optimization strategies}
\label{opt}
Finding a solution to \eqn{euler} is nontrivial. A stable optimization method is found by recasting \eqn{euler} in terms of $\psi(\br)=\sqrt{\rho(\br)}$\cite{PhysRevB.60.16350},
\begin{equation}\label{euler2}
\frac{\delta E[\psi^2]}{\delta \psi(\br)}-2 \mu \psi(\br)=0,
\end{equation}
in this way, there is no need to impose the constraint, $\rho(\br)>0$.

The algorithms employed to carry out the optimization have a long history and in many respects, they determine the computational efficiency of the entire OFDFT code. In DFTpy, we follow the common prescription. Given an initial $\psi(\br)$, the following steps are repeated until convergence is reached:
\begin{enumerate}
	\item Obtain the search direction vector $p_{k}(\br)$ with an optimization method of choice (e.g., conjugated gradient).
	\item Find an acceptable step size $\lambda_{k}$ along the vector $p_{k}(\br)$ using a line search strategy.
	\item Generate a new $\psi_{k+1}(\br)$ from $\psi_{k}(\br)$, $\lambda_{k}$ and $p_{k}(\br)$.
\end{enumerate}

For step (1), three main types of optimization methods are implemented in DFTpy: nonlinear conjugate gradient (CG)\cite{hestenes1952methods, fletcher1964function, polak1969note, polyak1969conjugate, fletcher1980practical,liu1991efficient, dai1999nonlinear}, limited memory Broyden-Fletcher-Goldfarb-Shanno (L-BFGS)\cite{liu1989limited} and truncated Newton (TN) methods\cite{nocedal2006numerical}. We tested the TN method to be the fastest method in DFTpy for most systems. However, in many instances (e.g. isolated systems), the TN method incurs into a high failure rate. Because in L-BFGS there is a need to store the last several updates of $ \psi $ and gradient, the memory cost is larger than for other methods. CG, instead, is the most stable among these methods, with several available options for updating $p_{k}$. In DFTpy, line search can be performed by the algorithms in {\tt SciPy.Optimize}. 

There are two ways to carry out an optimization: one is direct minimization of the energy functional, and another is the optimization of the residual [i.e., the result of the evaluation of \eqn{euler2}]. The optimizing function, $ \psi_{k+1} $, can be updated by $ \psi_{k+1}=\psi_{k}+\lambda_{k}p_{k} $, then normalized to $ N_{e} $. However, such a scaling scheme is not always stable. An alternative approach is to use an orthogonalization scheme prescribing $ p_{k} $ to be orthogonal to $ \psi_{k} $ and normalized to $ N_{e} $. The update can take the form \cite{jiang2004conjugate} $ \psi_{k+1}=\psi_{k}\cos(\lambda_{k})+p_{k}\sin(\lambda_{k}) $.

For those systems with inhomogeneous electron densities (such as clusters), convergence is very slow and can be very time consuming. For this reason, in DFTpy we implemented a multi-step density optimization scheme. In this scheme, the number of grid points needed to represent the electron density are determined dynamically and typically increase together with the optimization steps. We start out by carrying out a full density optimization on a coarse grid and then we interpolate the converged  density onto a finer grid leading to substantial savings. For example, if the grid spacing of the coarse grid is twice larger than the finer grid, the timing is decreased by $1/8$. For this scheme, the bigger the density inhomogeneities in the ground state density, the greater the efficiency improvement. In the next section, we will present an analysis of the timings and overall computational savings yielded by the new multi-step optimization method.

\subsection{Leveraging existing techniques}
DFTpy leverages fast algorithms, such as FFTs for Fourier transforms \cite{FFTW05}, and Particle-Mesh Ewald (PME) scheme for the computation of ionic structure factors \cite{Essmann,choly2003fast,hung2009accurate}. These are the most time consuming operations when large scale simulations are targeted \cite{PROFESS1.0,shao2016n}.

For FFTs, DFTpy encodes two modules: {\tt Numpy.fft} and {pyFFTW} \cite{pyfftw}. While {\tt Numpy.fft} is a portable FFT implementation, {pyFFTW} is perhaps the most efficient FFT under a {Python} environment that shares the same interface of {\tt Numpy.fft}. As FFTs are one of the most time consuming operations, it is worth to further improve them. For example, {Reikna} \cite{reikna} seems to offer a better interface to {PyCUDA} (the Python APIs for CUDA software to run on GPUs) compared to {pyFFTW}. Additionally, Google's {TensorFlow} \cite{tensorflow2015-whitepaper} also provides a GPU enabled FFT implementation ({\it via} Cuda FFT). Thus, in future DFTpy releases, we will develop APIs to both  {Reikna} and  {TensorFlow} modules.

Regarding the computation of the ionic structure factor, when a large number of ions is considered in the simulation (e.g., more than 1000 ions), the vanilla $O(N^2)$ method is no longer viable and, instead, the PME method is commonly employed. To our knowledge, there are no tested, open-source {Python} modules for PME. Thus, DFTpy has an in-house PME implementation, taking advantage of {SciPy} methods when possible. However, this may change in future releases if such a PME {Python} module had to become available.

%% file: applications.tex
Throughout this section, the calculations are carried out with the bulk-derived local pseudopotentials \cite{huang2008} (BLPS) and optimal effective local pseudopotentials (OEPP) \cite{mi2016first}, and the LDA xc functional parametrization by Perdew and Zunger \cite{Perdew1981}. Timing tests are performed starting from a face-centered cubic (fcc) Aluminum crystal with a lattice constant of 4.05 \r{A}, and a kinetic energy cutoff of 600 eV. This is sufficient to converge the total energy to below 1 meV/atom. 

\subsection{Optimization of the electron density}
\begin{figure}[htp]
	\centering
	\setlength{\labelsep}{-1em}
	\sidesubfloat[]{\includegraphics[width=0.45\textwidth]{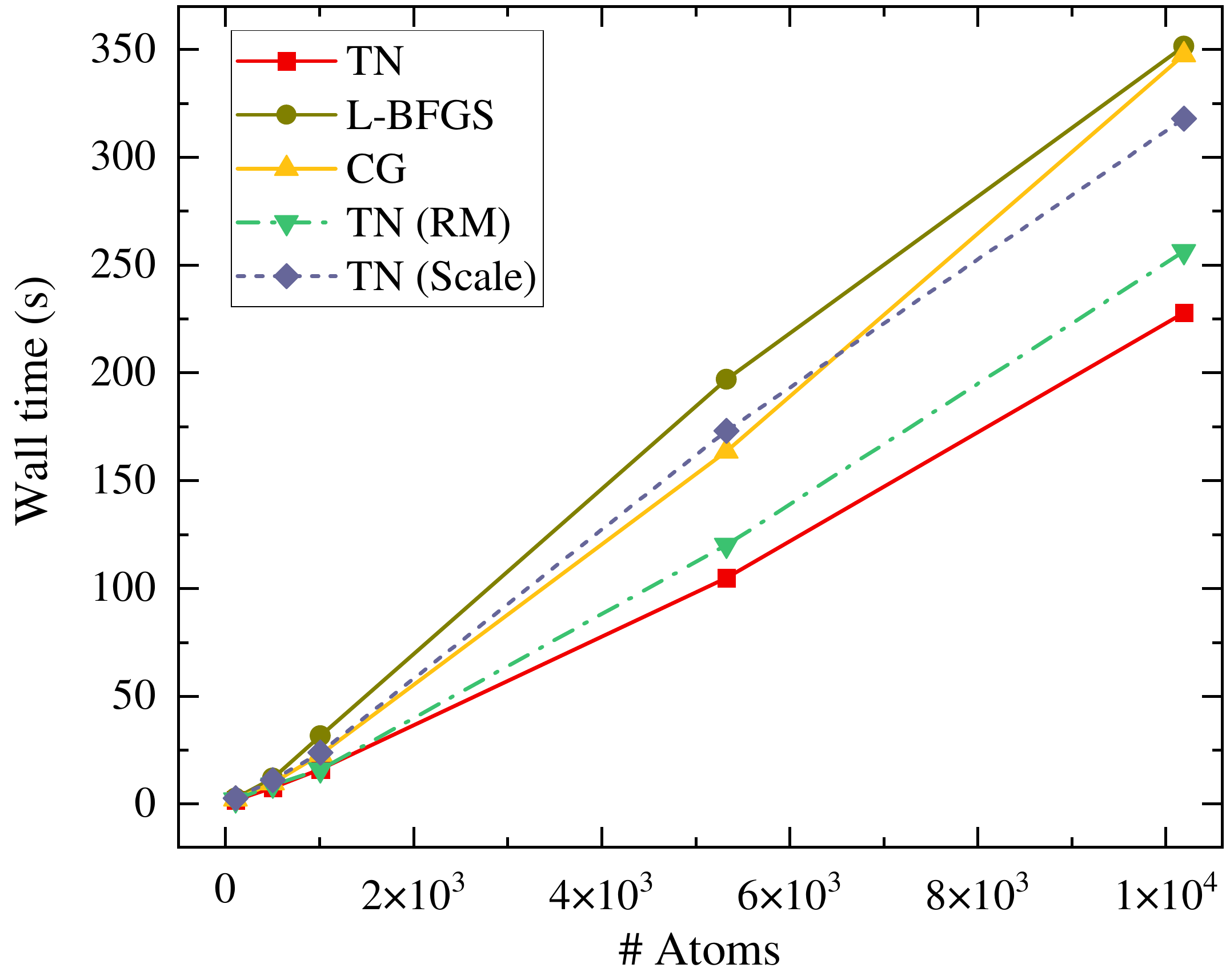}\label{fig:opt_a}} \hspace{1em}
	\sidesubfloat[]{\includegraphics[width=0.45\textwidth]{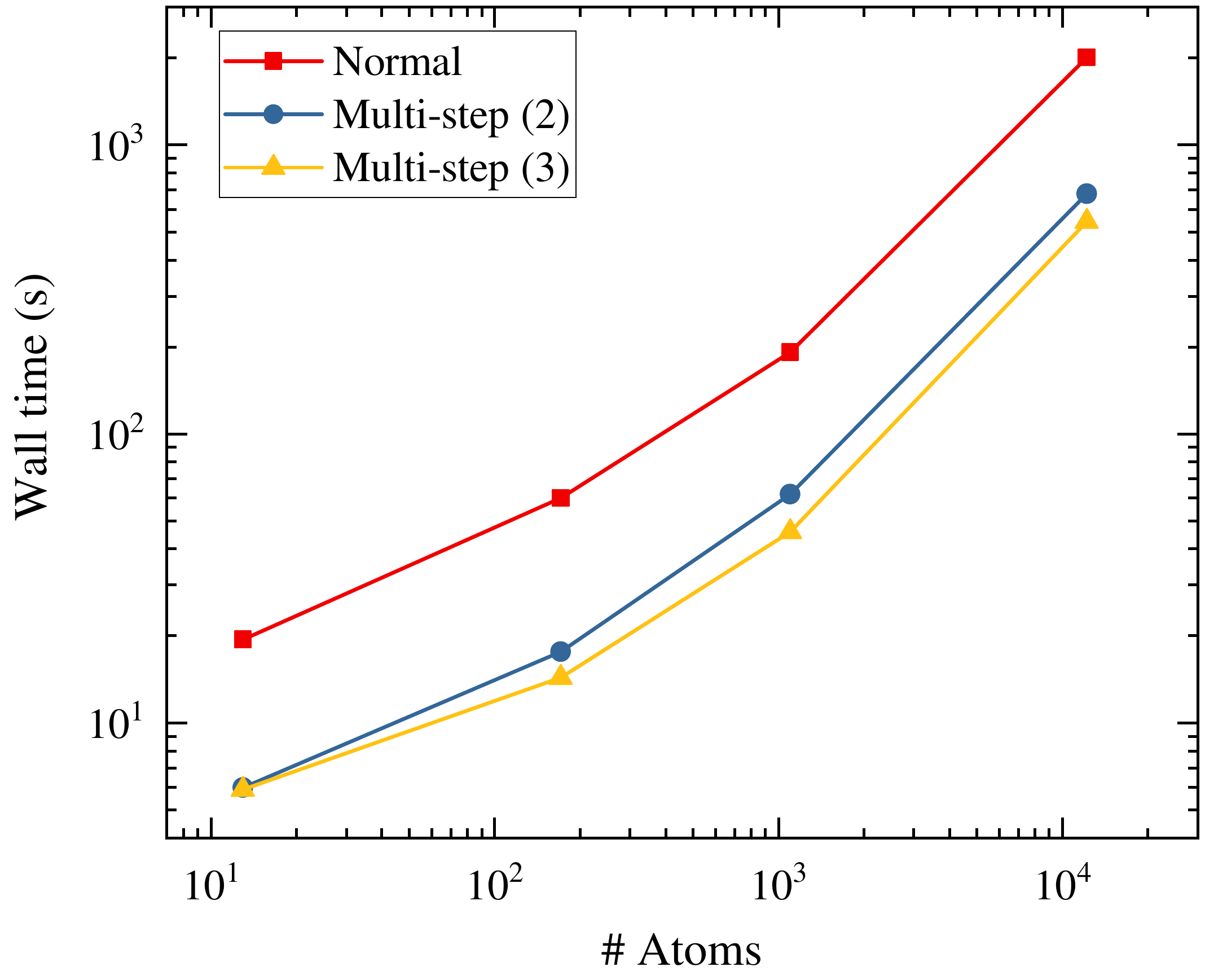}\label{fig:opt_b}}
	\caption{\label{fig:opt} Timings (wall time) for density optimizations carried out with different optimization methods. (a) Comparing optimizers on bulk Al supercells. (b) Comparing two- and three-step optimization to a vanilla density optimization for Al clusters.}
\end{figure}

Figure \ref{fig:opt}\subref{fig:opt_a} shows the total wall times required for the electron density optimization of systems containing up to $\sim$10,000 Al atoms using several optimization methods: CG, TN (regular, residual minimization and scaled, i.e., normalizing the density to the number of electrons), and L-BFGS. All methods show an approximate linear scaling execution time with system size. TN performs better than L-BFGS and CG methods. The residual minimization (RM) scheme with TN method, also presented in the figure, performs comparably to the energy minimization, and the scaling scheme shows good performance. We conclude that TN provides the most efficient optimization. Thus, TN is adopted for all the following calculations of bulk systems. 

The performance of the multi-step density optimization scheme described in \ref{opt} in comparison to a vanilla density optimization of Al clusters is shown in inset \subref{fig:opt_b} of Figure \ref{fig:opt}. In the calculation, we used the same structures as in Table \ref{tlmgp}, and for KEDF we use $ T_{\rm TF}+T_{\rm vW} $. For each step, CG is found to be more stable than TN method for isolated systems and is employed in the density optimization. The results show that the multi-step scheme speeds up the calculation by a factor of 2, demonstrating the high-efficiency of this multi-step scheme. In particular, a two-step scheme already brings most of the achievable savings, and a three-step scheme further improves, even though by a much smaller margin.

\subsection{Linear scalability up to one million atoms}
OFDFT methods are developed because they hold the promise to be able to describe realistically sized systems. In materials science, typical system sizes considered by the experiments involve thousands to well over millions of atoms. Will KSDFT ever be able to approach such systems? While it is hard to make a prediction at this particular point in history with quantum computing and machine learning spearheading new and potentially disruptive avenues of exploration, it is clear that current KSDFT algorithms (with exception of divide and conquer methods leveraging a mixture of KSDFT and OFDFT such as subsystem DFT \cite{krishtal2015subsystem}) and software are far from  being able to approach million-atom system sizes. OFDFT is developed to precisely fill this gap\cite{Hung2009163,shao2018large}.

DFTpy enters this playing field with an essentially single-core implementation (possibly enhanced by multithreading from OpenMP implementations of some underlying modules which are, however, not employed in this work). We stress here that a single core is perhaps all that is needed when system sizes of such dimensions are approached. This is because the complexity of sampling becomes a true computational bottleneck. Several thousands or even millions of structures need to be sampled in large-scale simulations, which make farming-type parallelization more efficient than single executions of parallel codes.

\begin{figure}[htp]
	\centering
	\includegraphics[width=0.8\textwidth]{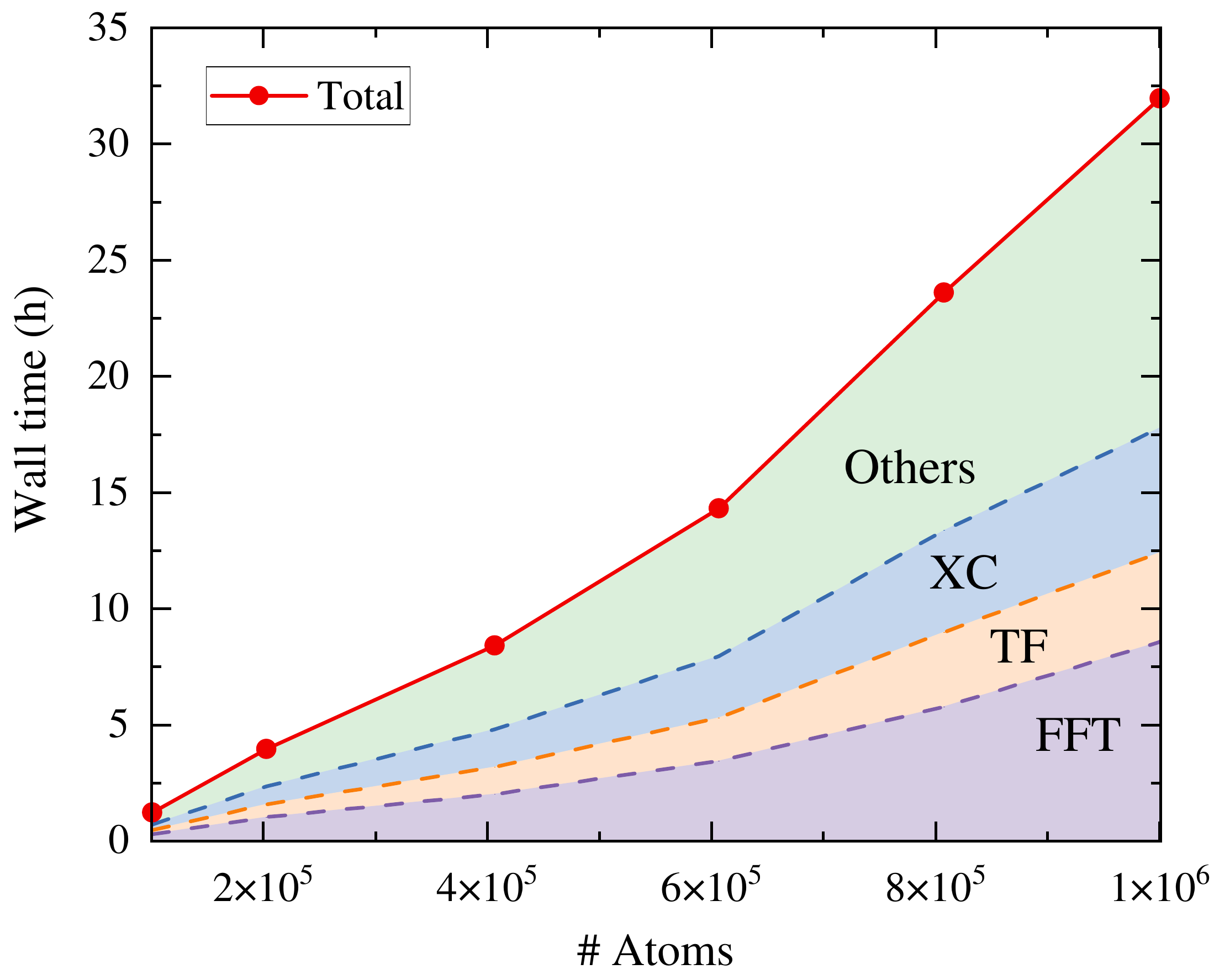}
	\caption{\label{fig:million} Timings (wall time) for density optimization on fcc aluminum for systems up to one million atoms with truncated Newton method.}
\end{figure}

To our knowledge, the largest system size ever approached by single-processor OFDFT software is 13,500 atoms \cite{PROFESS1.0}. At the same time, the largest system ever approached by parallel OFDFT codes reached $\sim$4 million atoms using 2,048 processors\cite{shao2016n}. To test the computational usefulness and efficiency of DFTpy, we perform a density optimization on the fcc Al supercell up to 1,000,188 atoms with a single processor. The total time and time-per-call for the total potential as a function of the number of atoms are presented in Figure \ref{fig:million}. From the figure, we can see that DFTpy still shows approximately linear scaling behavior with the number of atoms even for the large systems considered. The total time for simulating the $\sim$1 million atom system on a single core is only $\sim$32 h and can be further reduced to $\sim$20 h by employing the slightly lower cutoff of 500 eV which can still converge the total energy to within 1 meV/atom. We also notice that the FFT only accounts for $\sim$25\% of the total time, and surprisingly the time cost of $T_{\rm TF}$ and LDA exchange–correlation are comparable to the FFT. 

Thus, even though Python brings many important qualities to the developed software, it also poses few headaches. The example just mentioned shows that operations as simple as the power (i.e., $a=b^c$, involved in the evaluation of LDA functionals such as $T_{TF}$ and Dirac's exchange) can be inefficient when NumPy is used. Even though this comes at a linear cost, the prefactor is substantial, making the evaluation of the Thomas-Fermi functional much too expensive as seen in the figure. This issue will be tackled in future releases of DFTpy, for example by employing Pythran \cite{guelton2015pythran} or low-level languages for such operations.

\subsection{DFTpy+ASE: Dynamics of liquid aluminum}
Molecular dynamics (MD) is a widely used simulation technique in materials science and chemistry, useful to study structural and dynamic properties of materials. It is quite straightforward to develop an API that combines DFTpy and ASE to perform MD simulations. 

To showcase this API, we target a known success story for OFDFT. That is, the simulation of structure and dynamics of liquid metals, and particularly liquid Al. In Table \ref{tab:bulk}, we first show that DFTpy with the Wang-Teter (WT) functional is capable of predicting the correct equation of state for bulk Al. The equilibrium bulk structure is found numerically as well as via optimization (again, carried out via DFTpy+ASE) which agree with the fitted results from total energy values.

\begin{table}[htp]
	\caption{Bulk properties of fcc Al calculated by KSDFT and OFDFT methods. $ V_{0} $ is the equilibrium volume (\r{A}$^{3}$/atom), $E_{0}$ is the total energy (eV/atom), and $B_{0}$ is the bulk modulus (GPa). ``Relaxation'' refers to values obtained by OFDFT after a structure relaxation using DFTpy+ASE. }
	\begin{tabular}{cccc}
		\hline
		& $ V_{0} $        & $ E_{0} $        & $ B_{0} $       \\
		\hline
		KSDFT      & 15.644 & $-$57.951 & 82.94 \\
		OFDFT      & 15.819 & $-$57.934 & 84.97 \\
		Relaxation & 15.821 & $-$57.934 & $--$   \\
		\hline   
	\end{tabular}
\label{tab:bulk}
\end{table}

\begin{figure}[h]
	\centering
	\includegraphics[width=0.8\textwidth]{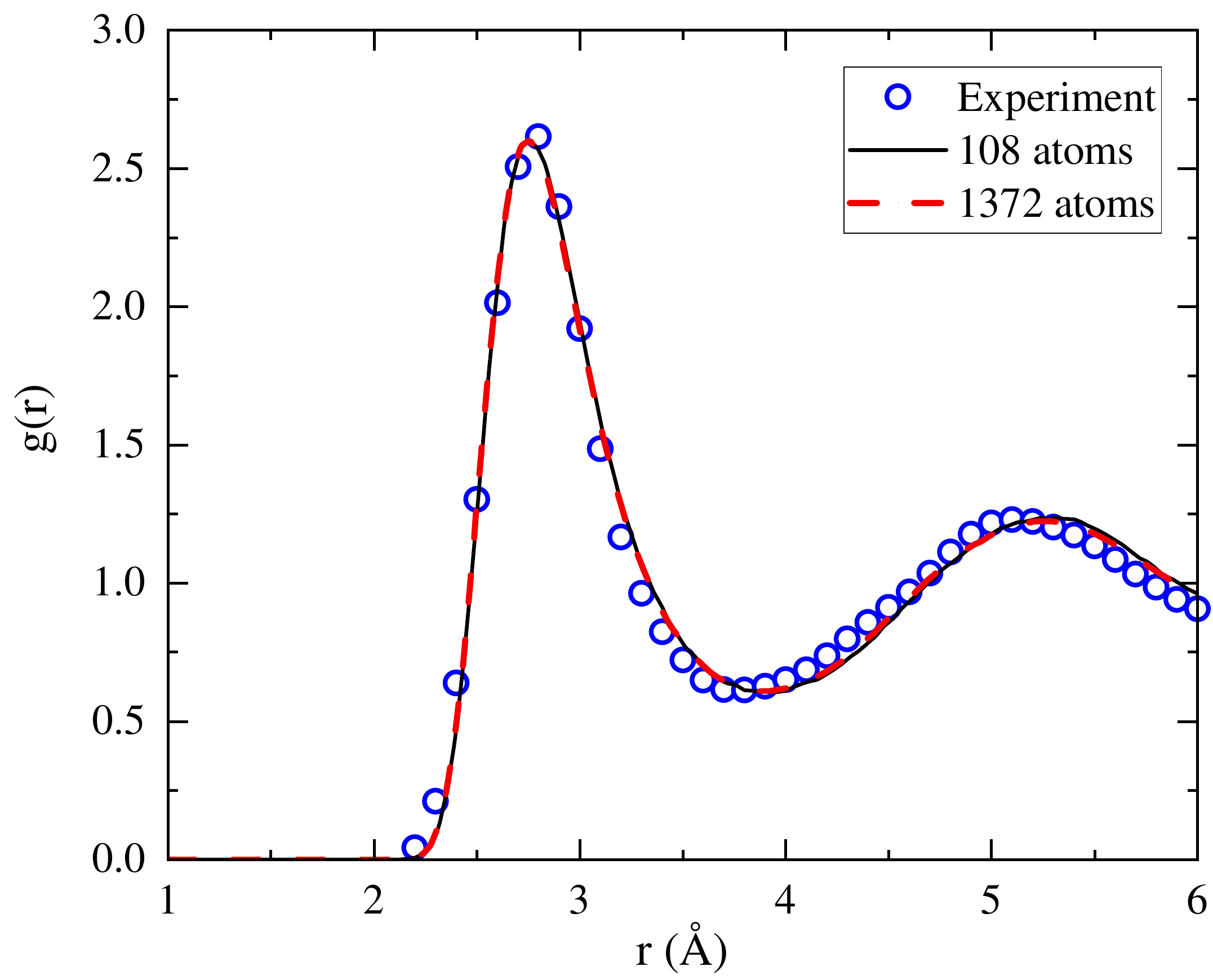}
	\caption{\label{fig:md}Pair distribution functions $g(r)$ for liquid Al at experimental conditions compared to X-ray diffraction data \cite{waseda1980structure}.}
\end{figure}

We then proceeded to carry out MD simulations in the canonical ensemble (NVT) for liquid Al at the experimental density 2.35 g/cm$ ^3 $ and the temperature of 1023 K\cite{waseda1980structure}. We first consider a small system size of 108 atoms and then we also tackle a 1,372 atom system. The time step used is 2 fs, and a Langevin thermostat\cite{allen2017computer} is used. Except a uniform density as the initial guess density in the first step, the initial density is given by optimization density of previous step in following steps, which further reduces the wall time. Figure \ref{fig:md} shows that our simulation results are in very good agreement with experimental data. DFTpy simulates the 108 atoms for 20,000 steps in only 37368 s ($\sim$10 h). To study finite-size effects on the $ g(r) $, we also carried out a simulation with a larger cell containing 1,372 atoms. The results in the figure show that finite-size effects are negligible for this system. Here, $ g(r) $ were averaged over 10,000 steps after equilibration.

%% file: ease.tex

\subsection{Time-Dependent OFDFT}


The hydrodynamic approach to time-dependent DFT (TD-DFT) has shown great promise for understanding plasmonics \cite{Cirac__2016,Banerjee_2008}, and the response of bulk metals \cite{PhysRevLett.111.175002}, metal surfaces \cite{Bennett_1970,Liebsch_1997}, and metal clusters \cite{Banerjee_2000}. Its applications, however, have been limited to model systems, such as jellium \cite{Liebsch_1997}, jellium spheres \cite{Banerjee_2000}, and other models \cite{Cirac__2016}. Even though these models are useful, as they provide a qualitative picture of the physics, a predictive and quantitative model can only be achieved when the atomistic details of the systems are taken into account. This is exactly our aim in this new implementation in DFTpy. Thus, in this section we present an implementation of atomistic hydrodynamic TD-DFT which we call TD-OFDFT, hereafter.

The theory follows closely OFDFT \cite{PhysRevLett.111.175002,Harbola_1998}, and introduces a ``collective orbital'' $\psi(\br)$, where $|\psi(\br)|^2=\rho(\br)$. We then solve the associated Schr\"{o}dinger-like equation.  Namely,
\begin{equation}
	\label{eq:of-ham}
\hat{H}\psi({\br})=\mu\psi({\br}),
\end{equation}
where
\begin{equation}
\label{heq}
\hat{H} = -\frac{1}{2}\nabla^2+\frac{\delta{T\s\Pauli}}{\delta\rho(\br)} + v\s(\br).
\end{equation}
The Laplacian term comes from the minimization of the \vv\ (vW) term, $T\s\VW$. $T\s\Pauli=T\s-T\s\VW$ is the remaining part of the non-interacting kinetic energy and is included in the TD-DFT effective potential.

A similar approach can be formulated for the time dependent extension requiring the current density $\bj(\br,t)=\rho(\br,t)\nabla S(\br,t)$, where $S(\br,t)$ is a scalar velocity field.
Thus, we write the time-dependent collective orbital in the form of $\psi(\br,t)\equiv\sqrt{\rho(\br,t)}e^{iS(\br,t)}$, and then solve a time-dependent Schr\"{o}dinger-like equation,
\begin{equation}
i\frac{\partial\psi({\br,t})}{\partial t}=\hat{H}\psi({\br, t}).
\end{equation}
Following \eqn{heq}, the Hamiltonian in the above equation has the form,
\begin{equation}
\hat{H} = -\frac{1}{2}\nabla^2+\frac{\delta{T\s\Pauli}}{\delta\rho(\br,t)} + v\s(\br,t),
\end{equation}
which we implement in the adiabatic LDA (ALDA) approximation.

This formalism can be exploited in several flavors: real-time propagations \cite{PhysRevLett.111.175002}, and perturbatively \cite{Banerjee_2002}. In this work, we choose the former, as described in the following section.

\subsubsection{Implementation of real-time TD-OFDFT}

We implemented a Crank-Nicolson propagator with predictor-corrector to any desired order. The relevant equation to solve for this implicit propagator is \cite{cast2004},

\begin{equation}
\label{cn}
\left(1-i\frac{\diff t}{2}\hat{H}\right)\psi(t+\diff t)=\left(1+i\frac{\diff t}{2}\hat{H}\right)\psi(t).
\end{equation}

The real-time TD-OFDFT simulation follows the workflow:
\begin{enumerate}
	\item Optimize the ground state density.
	\item Build the Hamiltonian in \eqn{heq}.
	\item Apply an external perturbation to displace the system from the ground electronic state ({\it vide infra}).
	\item Calculate the new potential with the density $\rho(t_i)$ and update the Hamiltonian with the new potential.
	\item Solve the linear system in \eqn{cn}, and propagate 1 time step from $t_i$ to $t_{i+1}$ (including the predictor-corrector step).
	\item Compute dipole moment and current density.
	\item Loop steps 4-6 until the total propagation time is reached.
\end{enumerate}

A simple Jupyter notebook encoding the TD-OFDFT scheme is available in the {\it notebooks} section of the GitLab repository \cite{dftpygit}, as well as in the {\it tutorials} tab of DFTpy's manual \cite{dftpyweb}.

\subsubsection{Optical spectra of Mg$_8$ and Mg$_{50}$ clusters}

We choose Mg metal clusters as the systems of interest. The system is optimized to its ground state density $\rho_0(\br)$. At $t=0$, we introduce a laser kick with strength $k$ in the x-direction by setting the collective phase, $S(\br,t=0)=-ikx$, 
\begin{equation}
\psi({\br, t=0}) = \psi(\br)e^{-ikx},
\end{equation}
where $\psi(\br)=\sqrt{\rho_0(\br)}$.
We then propagate the system in real time and obtain the time-dependent dipole moment change
\begin{equation}
\delta\mu(t) = \int\br\bigg(\rho(\br,t)-\rho_0(\br)\bigg)\d\br.
\end{equation}
The oscillator strength is calculated using the following equation:
\begin{equation}
\sigma(\omega)=\omega\text{Im}\bigg[\delta\tilde{\mu}(\omega)\bigg].
\end{equation}

For simplicity, we employ the Thomas-Fermi-\vv\ functional \cite{weiz1935}, which was shown to perform well for finite, isolated systems such as the metal clusters considered in this work \cite{chan2001thomas}. We use the OEPP local pseudopotentials\cite{mi2016first} and the Perdew--Zunger LDA xc functional\cite{Perdew1981}. A kinetic energy cutoff of 400 eV and 850 eV was employed for Mg$_8$ and Mg$_{50}$, respectively. We indicate by TD-KSDFT the TD-DFT calculations carried out with the exact noninteracting kinetic energy (i.e., Kohn-Sham) which are performed with the embedded Quantum Espresso (eQE) code\cite{genova2017eqe}. 

\begin{figure}[h]
	\centering
	\includegraphics[width=0.70\textwidth]{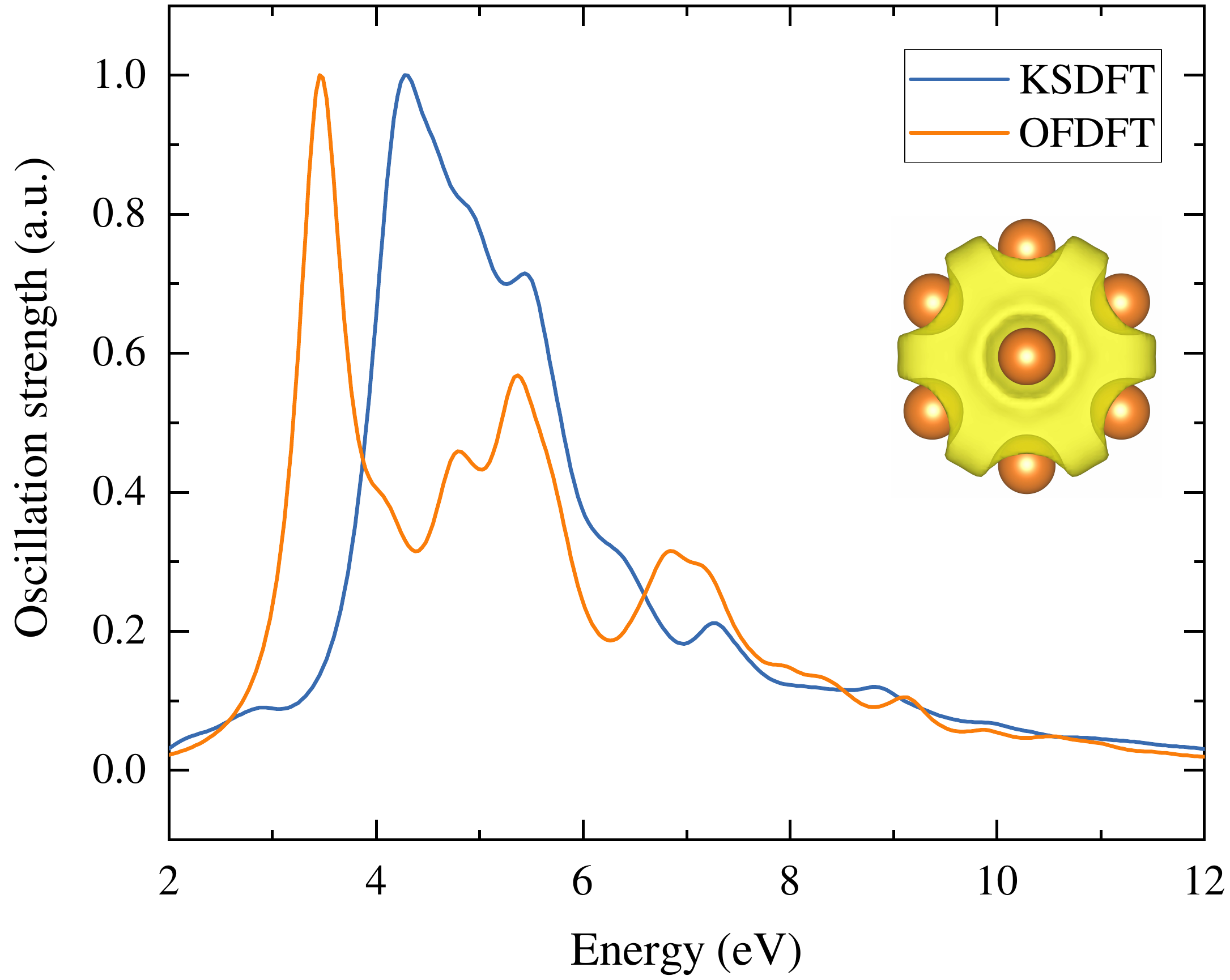}
	\caption{\label{fig:Mg8_spectra}Comparison of optical spectra obtained with  TD-OFDFT and TD-KSDFT for Mg$_8$. A view of the total density is given in the inset.}
\end{figure}

Metal clusters have been a common application of time-dependent Thomas-Fermi methods, including hydrodynamic OFDFT \cite{Harbola_1998}. The general consensus is that the larger the metal cluster, the closer the agreement with KSDFT. Banerjee and Harbola \cite{Banerjee_2000} showed that when OFDFT is applied to jellium spheres corresponding to cluster sizes of 100 atoms, the deviation between OFDFT and KSDFT in terms of the value of the static dipole polarizability goes below 20\%. For cluster sizes corresponding to 1000 atoms, the deviation goes below 2\%.

\begin{figure}[h]
	\centering
	\includegraphics[width=0.7\textwidth]{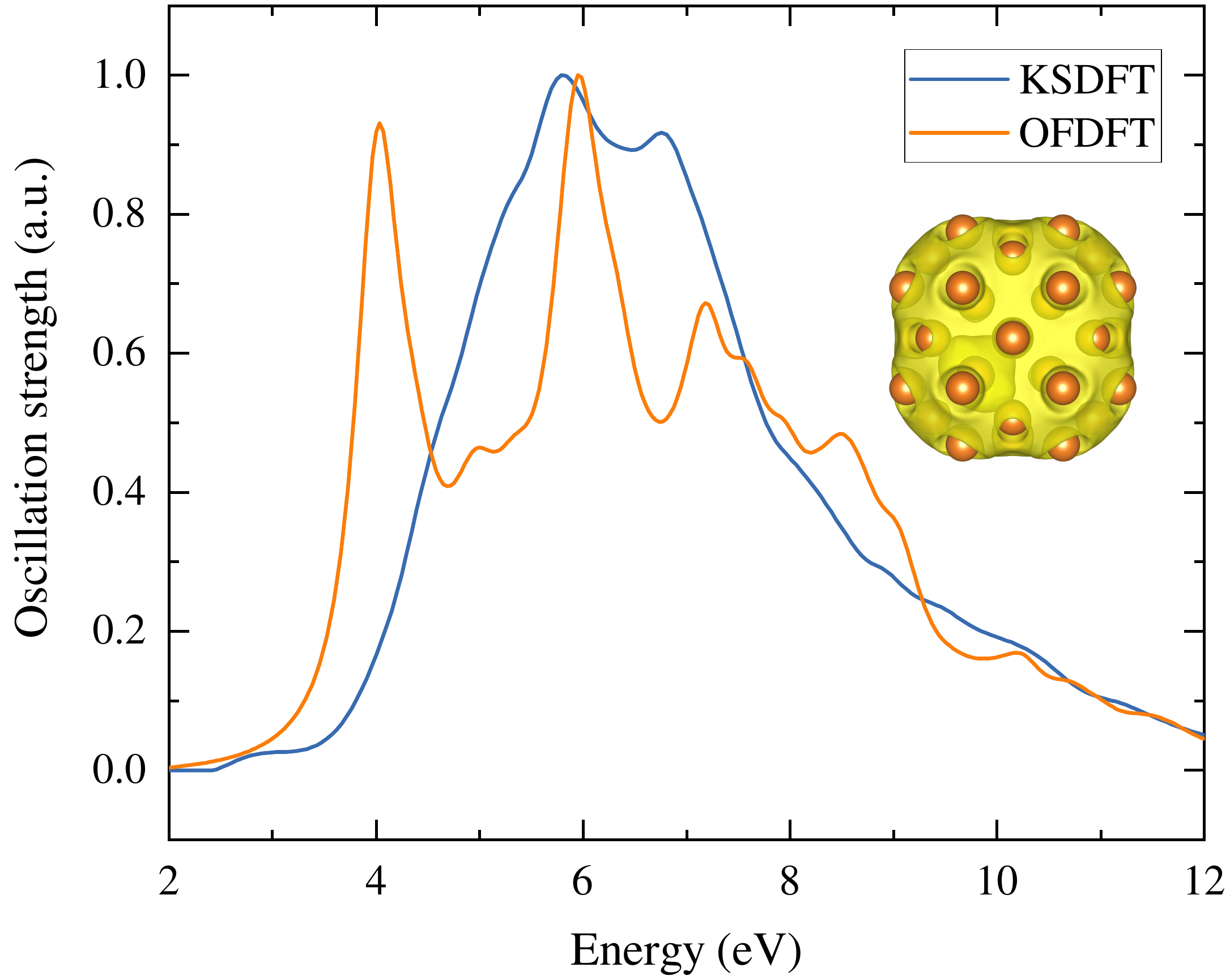}
	\caption{\label{fig:Mg50_spectra}Comparison of optical spectra computed with TD-OFDFT and TD-KSDFT for Mg$_{50}$. A view of the total density is given in the inset.}
\end{figure}

In a similar fashion, Figures \ref{fig:Mg8_spectra} and \ref{fig:Mg50_spectra}, show that our TD-OFDFT calculations yield spectra for Mg$_8$ and Mg$_{50}$ that are in fair agreement with TD-KSDFT.  

The agreement, however, is stronger in the Mg$_{50}$ syatem where the width and shape of the spectral envelope is better reproduced. The reason for such an agreement likely is the fact that Mg$_{50}$ can develop a uniform electron gas-like electronic structure in its core, a type of structure well characterized by a single orbital.

\subsubsection{Comparison of KS-DFT and OFDFT orbitals}
\begin{figure}[h]
	\centering
	\includegraphics[width=0.24\textwidth]{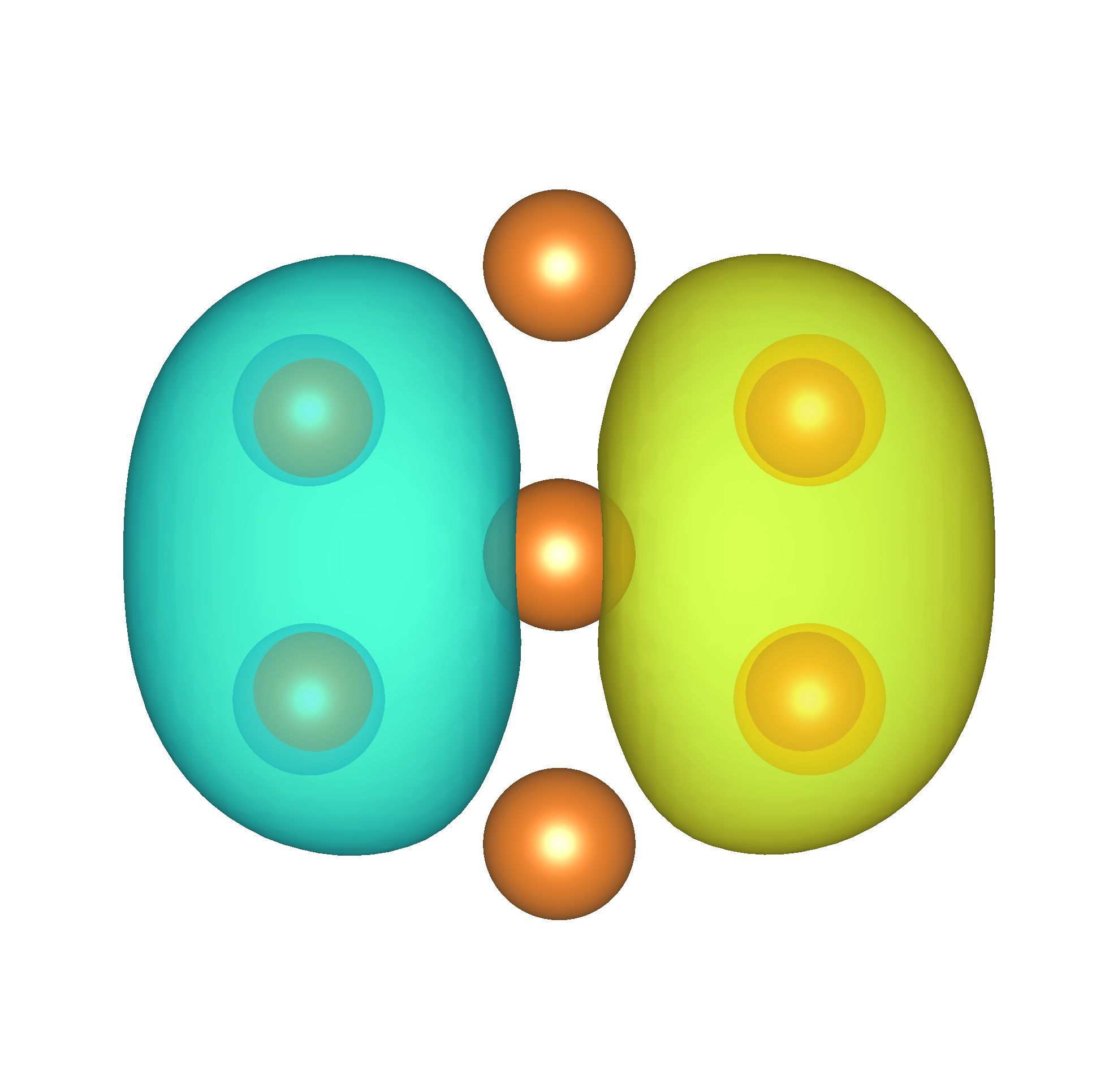}
	\includegraphics[width=0.24\textwidth]{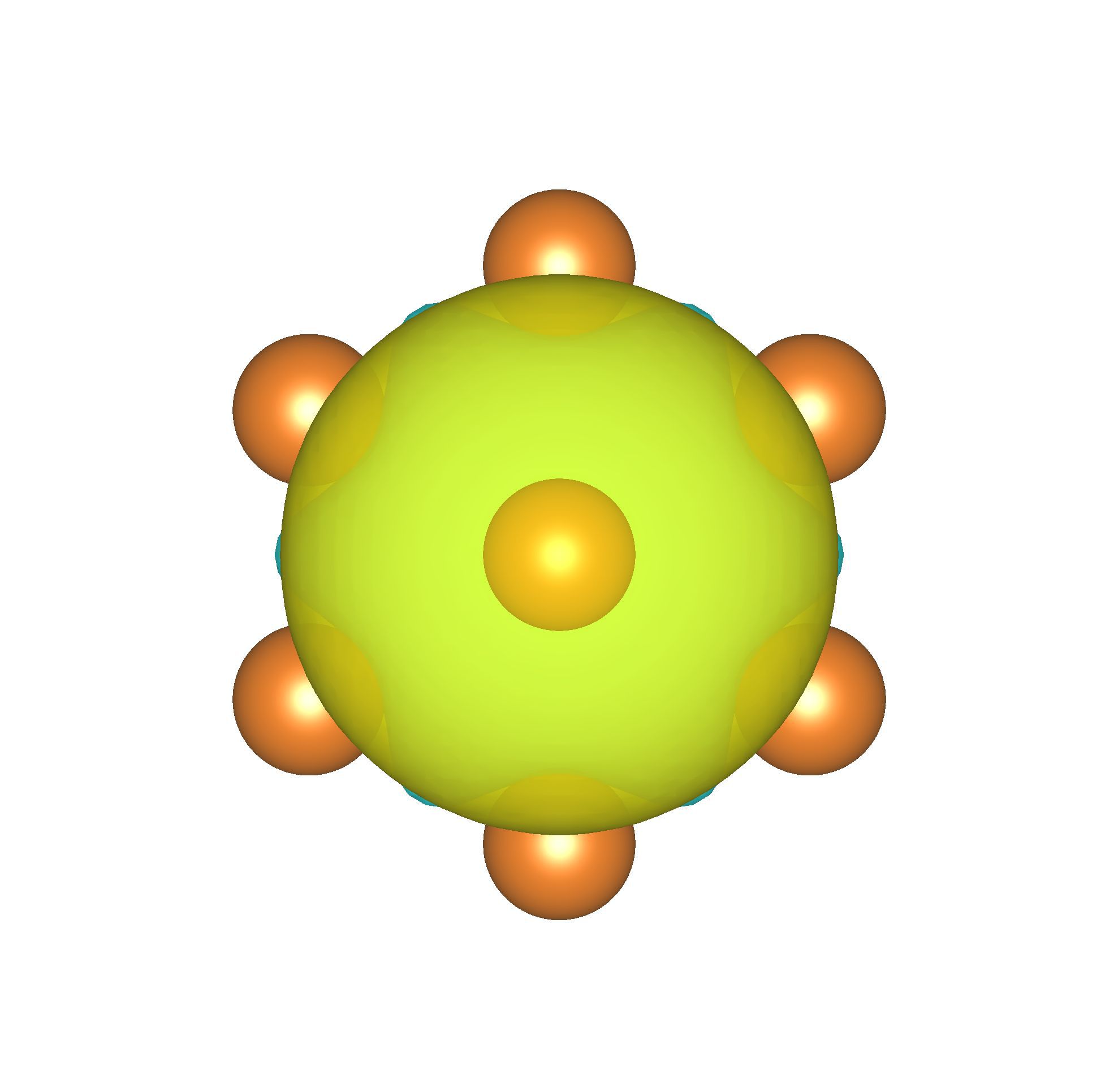}
	\includegraphics[width=0.24\textwidth]{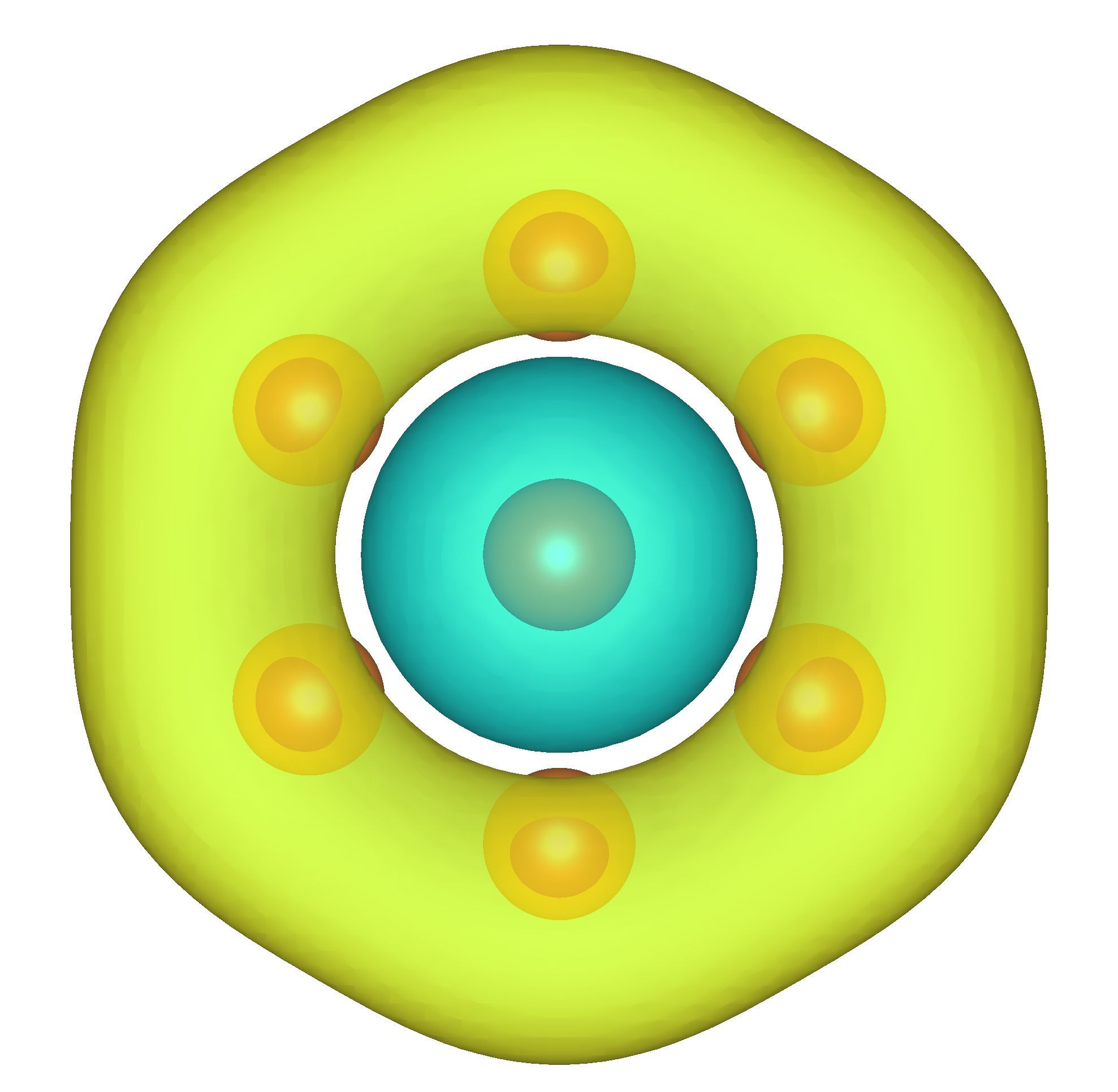}\par
	\includegraphics[width=0.24\textwidth]{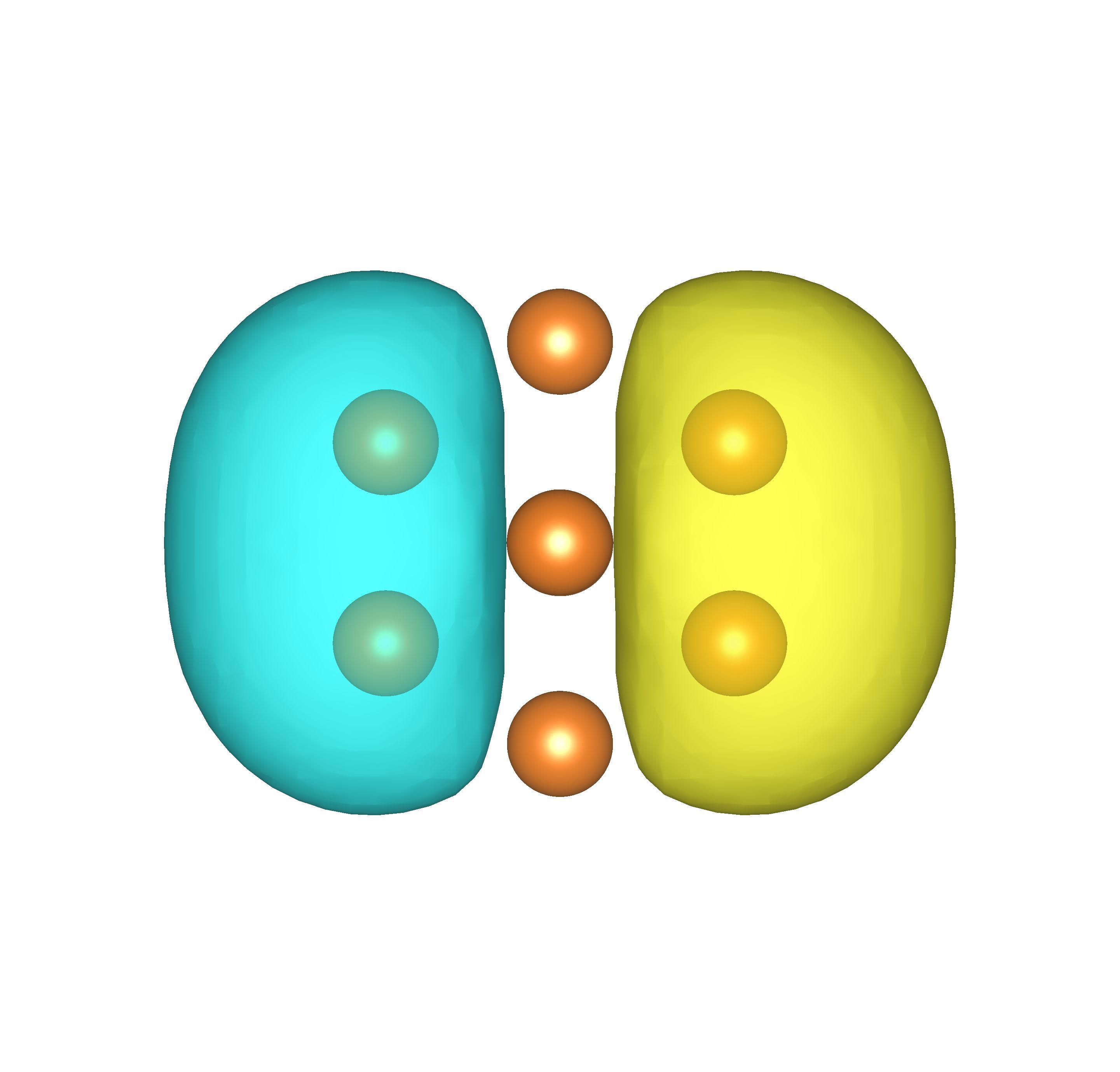}
	\includegraphics[width=0.24\textwidth]{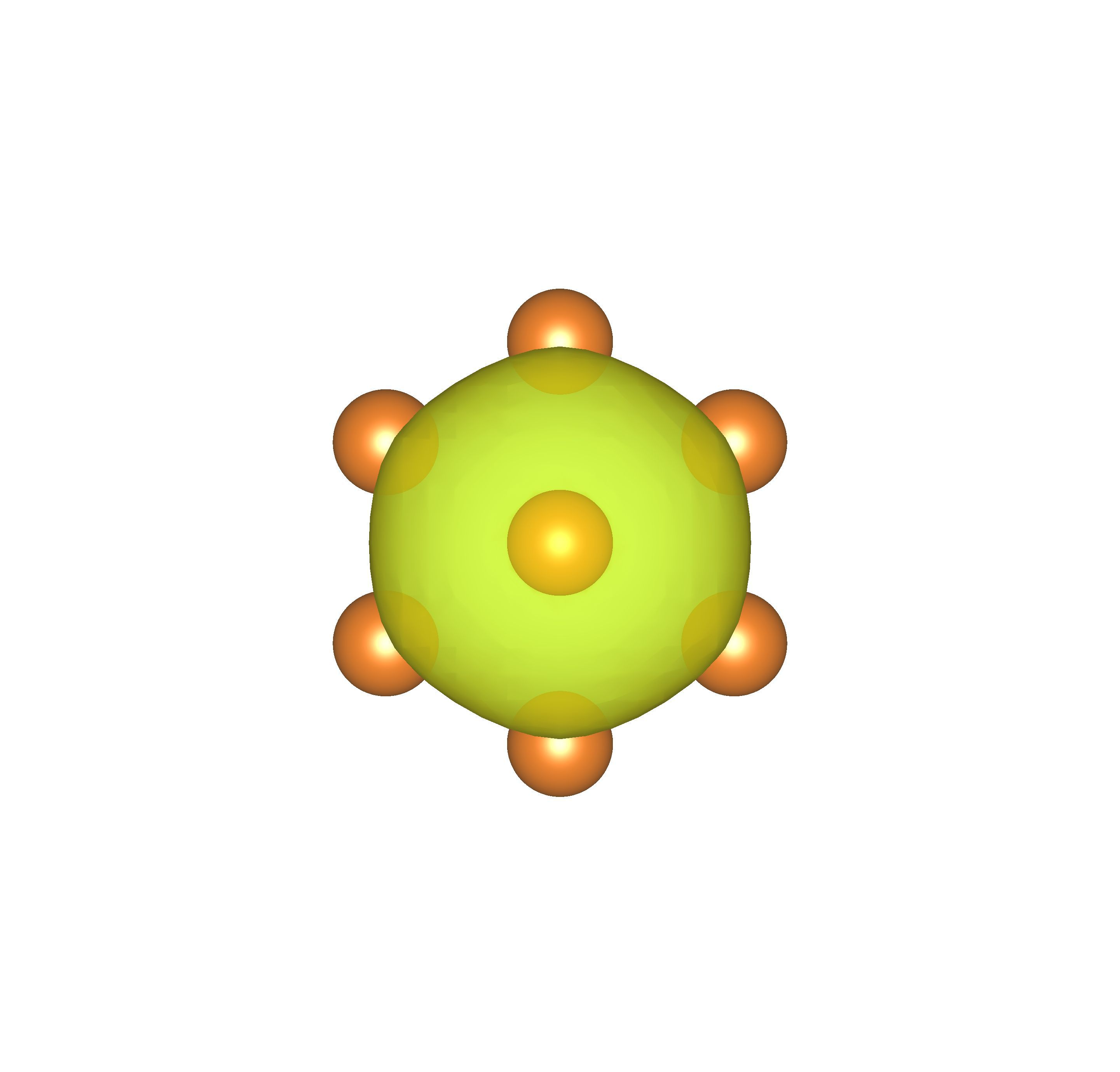}
	\includegraphics[width=0.24\textwidth]{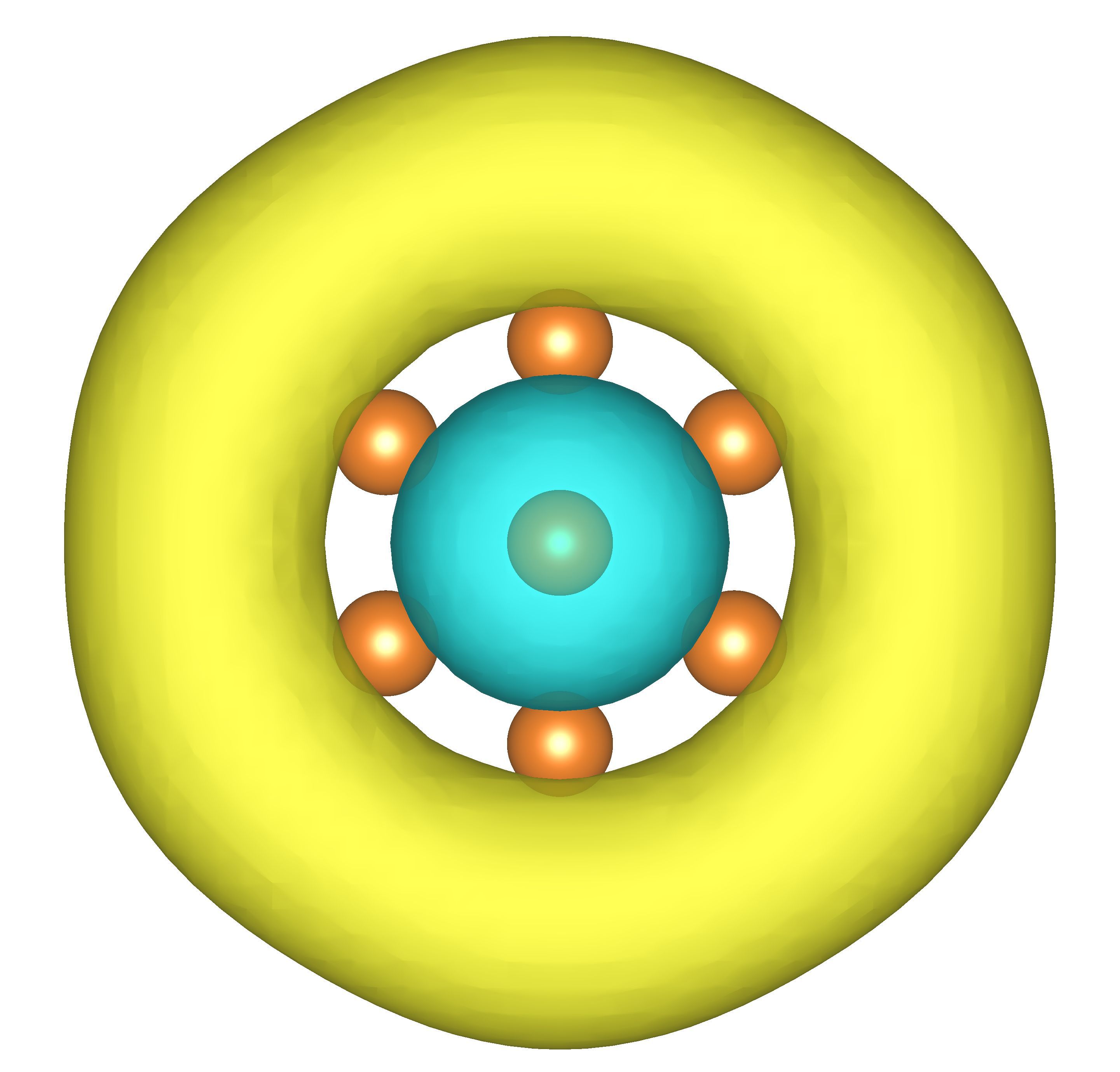}
	\caption{\label{orbs} Mg$_8$ KSDFT virtual orbitals: LUMO, LUMO+2, LUMO+6 (above). OFDFT ``collective'' virtual orbitals: LUMO, LUMO+3, LUMO+6 (below).} 
\end{figure}

An interesting question is whether the collective orbitals recovered by the solution of \eqn{eq:of-ham} resemble the KS orbitals. In principle, the collection of occupied and virtual KS orbitals form a complete basis, and so do the OFDFT collective orbitals. Thus, if we had to compare a large number of KS and collective orbitals, we would find that they span the exact same Hilbert space. For these reasons, we consider the Mg$_8$ system, and limit the comparison to the low-lying orbitals. Specifically, we compare orbitals within {5.0} eV from the Fermi energy, which corresponds to the first peak in the optical spectra. These comprise 17 OFDFT collective orbitals (1 occupied and 16 virtual) and 32 KS orbitals (8 occupied and 24 virtual). Three KSDFT and OFDFT virtuals are displayed in Figure \ref{orbs}. 

A direct comparison of OF and KS orbitals cannot be done visually. Therefore, we set up a rectangular overlap matrix, $S_{ij}=\langle \psi_i^\text{KS} | \psi_j^\text{OF}\rangle$ and compute its singular value decomposition. The distribution of the singular values are collected in Table \ref{orbcomp}.

%

\begin{table}[h]
	\centering
	\begin{tabular}{c c c}
		\hline
		&\multicolumn{2}{c}{Number of singular values}\\\cline{2-3}
		Range&Occ + Virt & Virt only\\\hline
		0.9 - 1.0 & 14 & 7 \\
		0.8 - 0.9 &  2 & 1 \\
		0.7 - 0.8 &  1 & 1 \\
		0.6 - 0.7 &  & 1 \\
		0.5 - 0.6 &  & 1 \\
		0.4 - 0.5 &	 & 1 \\
		0.3 - 0.4 &	 & 4 \\\hline
		
	\end{tabular}
	\caption{\label{orbcomp}Distribution of the singular values of the overlap matrix, $S_{ij}$, between KS and OF-DFT orbitals for Mg$_8$. A selection of virtuals can be inspected in Figure \ref{orbs}.}
\end{table}

If occupied and virtuals are included in the singular value decomposition,  the OF orbitals can be essentially exactly represented as a linear combination of the KS orbitals (the majority of the singular values are close to $1$). However, if only virtual orbitals within a {5.0} eV energy window from the Fermi energy are considered, the OF orbitals can only be partially decomposed into KS virtual orbitals. Thus, the comparison shows that OFDFT and KSDFT orbitals are similar if occupied and virtuals are compared. The virtual spaces, however, are only partially similar.

%% file: paper.bbl
\begin{thebibliography}{72}%
\makeatletter
\providecommand \@ifxundefined [1]{%
 \@ifx{#1\undefined}
}%
\providecommand \@ifnum [1]{%
 \ifnum #1\expandafter \@firstoftwo
 \else \expandafter \@secondoftwo
 \fi
}%
\providecommand \@ifx [1]{%
 \ifx #1\expandafter \@firstoftwo
 \else \expandafter \@secondoftwo
 \fi
}%
\providecommand \natexlab [1]{#1}%
\providecommand \enquote  [1]{``#1''}%
\providecommand \bibnamefont  [1]{#1}%
\providecommand \bibfnamefont [1]{#1}%
\providecommand \citenamefont [1]{#1}%
\providecommand \href@noop [0]{\@secondoftwo}%
\providecommand \href [0]{\begingroup \@sanitize@url \@href}%
\providecommand \@href[1]{\@@startlink{#1}\@@href}%
\providecommand \@@href[1]{\endgroup#1\@@endlink}%
\providecommand \@sanitize@url [0]{\catcode `\\12\catcode `\$12\catcode
  `\&12\catcode `\#12\catcode `\^12\catcode `\_12\catcode `\%12\relax}%
\providecommand \@@startlink[1]{}%
\providecommand \@@endlink[0]{}%
\providecommand \url  [0]{\begingroup\@sanitize@url \@url }%
\providecommand \@url [1]{\endgroup\@href {#1}{\urlprefix }}%
\providecommand \urlprefix  [0]{URL }%
\providecommand \Eprint [0]{\href }%
\providecommand \doibase [0]{http://dx.doi.org/}%
\providecommand \selectlanguage [0]{\@gobble}%
\providecommand \bibinfo  [0]{\@secondoftwo}%
\providecommand \bibfield  [0]{\@secondoftwo}%
\providecommand \translation [1]{[#1]}%
\providecommand \BibitemOpen [0]{}%
\providecommand \bibitemStop [0]{}%
\providecommand \bibitemNoStop [0]{.\EOS\space}%
\providecommand \EOS [0]{\spacefactor3000\relax}%
\providecommand \BibitemShut  [1]{\csname bibitem#1\endcsname}%
\let\auto@bib@innerbib\@empty
\bibitem [{\citenamefont {Witt}\ \emph {et~al.}(2018)\citenamefont {Witt},
  \citenamefont {Beatriz}, \citenamefont {Dieterich},\ and\ \citenamefont
  {Carter}}]{witt2018orbital}%
  \BibitemOpen
  \bibfield  {author} {\bibinfo {author} {\bibfnamefont {W.~C.}\ \bibnamefont
  {Witt}}, \bibinfo {author} {\bibfnamefont {G.}~\bibnamefont {Beatriz}},
  \bibinfo {author} {\bibfnamefont {J.~M.}\ \bibnamefont {Dieterich}}, \ and\
  \bibinfo {author} {\bibfnamefont {E.~A.}\ \bibnamefont {Carter}},\ }\bibfield
   {title} {\enquote {\bibinfo {title} {Orbital-free density functional theory
  for materials research},}\ }\href@noop {} {\bibfield  {journal} {\bibinfo
  {journal} {J. M. Res.}\ }\textbf {\bibinfo {volume} {33}},\ \bibinfo {pages}
  {777--795} (\bibinfo {year} {2018})}\BibitemShut {NoStop}%
\bibitem [{\citenamefont {Chen}\ \emph {et~al.}(2015)\citenamefont {Chen},
  \citenamefont {Xia}, \citenamefont {Huang}, \citenamefont {Dieterich},
  \citenamefont {Hung}, \citenamefont {Shin},\ and\ \citenamefont
  {Carter}}]{Chen2015228}%
  \BibitemOpen
  \bibfield  {author} {\bibinfo {author} {\bibfnamefont {M.}~\bibnamefont
  {Chen}}, \bibinfo {author} {\bibfnamefont {J.}~\bibnamefont {Xia}}, \bibinfo
  {author} {\bibfnamefont {C.}~\bibnamefont {Huang}}, \bibinfo {author}
  {\bibfnamefont {J.~M.}\ \bibnamefont {Dieterich}}, \bibinfo {author}
  {\bibfnamefont {L.}~\bibnamefont {Hung}}, \bibinfo {author} {\bibfnamefont
  {I.}~\bibnamefont {Shin}}, \ and\ \bibinfo {author} {\bibfnamefont {E.~A.}\
  \bibnamefont {Carter}},\ }\bibfield  {title} {\enquote {\bibinfo {title}
  {Introducing profess 3.0: An advanced program for orbital-free density
  functional theory molecular dynamics simulations},}\ }\href {\doibase
  http://dx.doi.org/10.1016/j.cpc.2014.12.021} {\bibfield  {journal} {\bibinfo
  {journal} {Computer Physics Communications}\ }\textbf {\bibinfo {volume}
  {190}},\ \bibinfo {pages} {228 -- 230} (\bibinfo {year} {2015})}\BibitemShut
  {NoStop}%
\bibitem [{\citenamefont {Shao}\ \emph {et~al.}(2018)\citenamefont {Shao},
  \citenamefont {Xu}, \citenamefont {Wang}, \citenamefont {Lv}, \citenamefont
  {Wang},\ and\ \citenamefont {Ma}}]{shao2018large}%
  \BibitemOpen
  \bibfield  {author} {\bibinfo {author} {\bibfnamefont {X.}~\bibnamefont
  {Shao}}, \bibinfo {author} {\bibfnamefont {Q.}~\bibnamefont {Xu}}, \bibinfo
  {author} {\bibfnamefont {S.}~\bibnamefont {Wang}}, \bibinfo {author}
  {\bibfnamefont {J.}~\bibnamefont {Lv}}, \bibinfo {author} {\bibfnamefont
  {Y.}~\bibnamefont {Wang}}, \ and\ \bibinfo {author} {\bibfnamefont
  {Y.}~\bibnamefont {Ma}},\ }\bibfield  {title} {\enquote {\bibinfo {title}
  {Large-scale ab initio simulations for periodic system},}\ }\href@noop {}
  {\bibfield  {journal} {\bibinfo  {journal} {Computer Physics Communications}\
  }\textbf {\bibinfo {volume} {233}},\ \bibinfo {pages} {78--83} (\bibinfo
  {year} {2018})}\BibitemShut {NoStop}%
\bibitem [{\citenamefont {Gavini}\ \emph {et~al.}(2007)\citenamefont {Gavini},
  \citenamefont {Knap}, \citenamefont {Bhattacharya},\ and\ \citenamefont
  {Ortiz}}]{gavini2007non}%
  \BibitemOpen
  \bibfield  {author} {\bibinfo {author} {\bibfnamefont {V.}~\bibnamefont
  {Gavini}}, \bibinfo {author} {\bibfnamefont {J.}~\bibnamefont {Knap}},
  \bibinfo {author} {\bibfnamefont {K.}~\bibnamefont {Bhattacharya}}, \ and\
  \bibinfo {author} {\bibfnamefont {M.}~\bibnamefont {Ortiz}},\ }\bibfield
  {title} {\enquote {\bibinfo {title} {Non-periodic finite-element formulation
  of orbital-free density functional theory},}\ }\href@noop {} {\bibfield
  {journal} {\bibinfo  {journal} {Journal of the Mechanics and Physics of
  Solids}\ }\textbf {\bibinfo {volume} {55}},\ \bibinfo {pages} {669--696}
  (\bibinfo {year} {2007})}\BibitemShut {NoStop}%
\bibitem [{\citenamefont {Chen}\ \emph {et~al.}(2016)\citenamefont {Chen},
  \citenamefont {Jiang}, \citenamefont {Zhuang}, \citenamefont {Wang},\ and\
  \citenamefont {Carter}}]{Chen_2016}%
  \BibitemOpen
  \bibfield  {author} {\bibinfo {author} {\bibfnamefont {M.}~\bibnamefont
  {Chen}}, \bibinfo {author} {\bibfnamefont {X.-W.}\ \bibnamefont {Jiang}},
  \bibinfo {author} {\bibfnamefont {H.}~\bibnamefont {Zhuang}}, \bibinfo
  {author} {\bibfnamefont {L.-W.}\ \bibnamefont {Wang}}, \ and\ \bibinfo
  {author} {\bibfnamefont {E.~A.}\ \bibnamefont {Carter}},\ }\bibfield  {title}
  {\enquote {\bibinfo {title} {Petascale orbital-free density functional theory
  enabled by small-box algorithms},}\ }\href {\doibase
  10.1021/acs.jctc.6b00326} {\bibfield  {journal} {\bibinfo  {journal} {Journal
  of Chemical Theory and Computation}\ }\textbf {\bibinfo {volume} {12}},\
  \bibinfo {pages} {2950--2963} (\bibinfo {year} {2016})}\BibitemShut {NoStop}%
\bibitem [{\citenamefont {Xia}\ and\ \citenamefont {Carter}(2012)}]{xia2012}%
  \BibitemOpen
  \bibfield  {author} {\bibinfo {author} {\bibfnamefont {J.}~\bibnamefont
  {Xia}}\ and\ \bibinfo {author} {\bibfnamefont {E.~A.}\ \bibnamefont
  {Carter}},\ }\bibfield  {title} {\enquote {\bibinfo {title}
  {{Density-decomposed orbital-free density functional theory for covalently
  bonded molecules and materials}},}\ }\href {\doibase
  10.1103/PhysRevB.86.235109} {\bibfield  {journal} {\bibinfo  {journal} {Phys.
  Rev. B}\ }\textbf {\bibinfo {volume} {86}},\ \bibinfo {pages} {235109}
  (\bibinfo {year} {2012})}\BibitemShut {NoStop}%
\bibitem [{\citenamefont {Xia}\ \emph {et~al.}(2012)\citenamefont {Xia},
  \citenamefont {Huang}, \citenamefont {Shin},\ and\ \citenamefont
  {Carter}}]{xia2012b}%
  \BibitemOpen
  \bibfield  {author} {\bibinfo {author} {\bibfnamefont {J.}~\bibnamefont
  {Xia}}, \bibinfo {author} {\bibfnamefont {C.}~\bibnamefont {Huang}}, \bibinfo
  {author} {\bibfnamefont {I.}~\bibnamefont {Shin}}, \ and\ \bibinfo {author}
  {\bibfnamefont {E.~A.}\ \bibnamefont {Carter}},\ }\bibfield  {title}
  {\enquote {\bibinfo {title} {{Can orbital-free density functional theory
  simulate molecules?}}}\ }\href {\doibase 10.1063/1.3685604} {\bibfield
  {journal} {\bibinfo  {journal} {J. Comp. Phys.}\ }\textbf {\bibinfo {volume}
  {136}},\ \bibinfo {pages} {084102} (\bibinfo {year} {2012})}\BibitemShut
  {NoStop}%
\bibitem [{\citenamefont {Mi}\ and\ \citenamefont
  {Pavanello}(2019)}]{mi2019LMGP}%
  \BibitemOpen
  \bibfield  {author} {\bibinfo {author} {\bibfnamefont {W.}~\bibnamefont
  {Mi}}\ and\ \bibinfo {author} {\bibfnamefont {M.}~\bibnamefont {Pavanello}},\
  }\bibfield  {title} {\enquote {\bibinfo {title} {Orbital-free dft correctly
  models quantum dots when asymptotics, nonlocality and nonhomogeneity are
  accounted for},}\ }\href@noop {} {\bibfield  {journal} {\bibinfo  {journal}
  {Phys. Rev. B Rapid Commun.}\ }\textbf {\bibinfo {volume} {100}},\ \bibinfo
  {pages} {041105} (\bibinfo {year} {2019})}\BibitemShut {NoStop}%
\bibitem [{\citenamefont {Xu}, \citenamefont {Wang},\ and\ \citenamefont
  {Ma}(2019)}]{Xu_2019}%
  \BibitemOpen
  \bibfield  {author} {\bibinfo {author} {\bibfnamefont {Q.}~\bibnamefont
  {Xu}}, \bibinfo {author} {\bibfnamefont {Y.}~\bibnamefont {Wang}}, \ and\
  \bibinfo {author} {\bibfnamefont {Y.}~\bibnamefont {Ma}},\ }\bibfield
  {title} {\enquote {\bibinfo {title} {Nonlocal kinetic energy density
  functional via line integrals and its application to orbital-free density
  functional theory},}\ }\href {https://doi.org/10.1103%2Fphysrevb.100.205132}
  {\bibfield  {journal} {\bibinfo  {journal} {Phy. Rev. B}\ }\textbf {\bibinfo
  {volume} {100}} (\bibinfo {year} {2019})}\BibitemShut {NoStop}%
\bibitem [{\citenamefont {Lehtomäki}\ \emph {et~al.}(2014)\citenamefont
  {Lehtomäki}, \citenamefont {Makkonen}, \citenamefont {Caro}, \citenamefont
  {Harju},\ and\ \citenamefont {Lopez-Acevedo}}]{Lehtom_ki_2014}%
  \BibitemOpen
  \bibfield  {author} {\bibinfo {author} {\bibfnamefont {J.}~\bibnamefont
  {Lehtomäki}}, \bibinfo {author} {\bibfnamefont {I.}~\bibnamefont
  {Makkonen}}, \bibinfo {author} {\bibfnamefont {M.~A.}\ \bibnamefont {Caro}},
  \bibinfo {author} {\bibfnamefont {A.}~\bibnamefont {Harju}}, \ and\ \bibinfo
  {author} {\bibfnamefont {O.}~\bibnamefont {Lopez-Acevedo}},\ }\bibfield
  {title} {\enquote {\bibinfo {title} {Orbital-free density functional theory
  implementation with the projector augmented-wave method},}\ }\href {\doibase
  10.1063/1.4903450} {\bibfield  {journal} {\bibinfo  {journal} {J. Chem.
  Phys.}\ }\textbf {\bibinfo {volume} {141}},\ \bibinfo {pages} {234102}
  (\bibinfo {year} {2014})}\BibitemShut {NoStop}%
\bibitem [{\citenamefont {Karasiev}, \citenamefont {Sjostrom},\ and\
  \citenamefont {Trickey}(2014)}]{Karasiev_2014}%
  \BibitemOpen
  \bibfield  {author} {\bibinfo {author} {\bibfnamefont {V.~V.}\ \bibnamefont
  {Karasiev}}, \bibinfo {author} {\bibfnamefont {T.}~\bibnamefont {Sjostrom}},
  \ and\ \bibinfo {author} {\bibfnamefont {S.}~\bibnamefont {Trickey}},\
  }\bibfield  {title} {\enquote {\bibinfo {title} {Finite-temperature
  orbital-free {DFT} molecular dynamics: Coupling profess and quantum
  espresso},}\ }\href {\doibase 10.1016/j.cpc.2014.08.023} {\bibfield
  {journal} {\bibinfo  {journal} {Computer Physics Communications}\ }\textbf
  {\bibinfo {volume} {185}},\ \bibinfo {pages} {3240--3249} (\bibinfo {year}
  {2014})}\BibitemShut {NoStop}%
\bibitem [{\citenamefont {Mi}\ \emph {et~al.}(2016{\natexlab{a}})\citenamefont
  {Mi}, \citenamefont {Shao}, \citenamefont {Su}, \citenamefont {Zhou},
  \citenamefont {Zhang}, \citenamefont {Li}, \citenamefont {Wang},
  \citenamefont {Zhang}, \citenamefont {Miao}, \citenamefont {Wang},\ and\
  \citenamefont {Ma}}]{ATLAS}%
  \BibitemOpen
  \bibfield  {author} {\bibinfo {author} {\bibfnamefont {W.}~\bibnamefont
  {Mi}}, \bibinfo {author} {\bibfnamefont {X.}~\bibnamefont {Shao}}, \bibinfo
  {author} {\bibfnamefont {C.}~\bibnamefont {Su}}, \bibinfo {author}
  {\bibfnamefont {Y.}~\bibnamefont {Zhou}}, \bibinfo {author} {\bibfnamefont
  {S.}~\bibnamefont {Zhang}}, \bibinfo {author} {\bibfnamefont
  {Q.}~\bibnamefont {Li}}, \bibinfo {author} {\bibfnamefont {H.}~\bibnamefont
  {Wang}}, \bibinfo {author} {\bibfnamefont {L.}~\bibnamefont {Zhang}},
  \bibinfo {author} {\bibfnamefont {M.}~\bibnamefont {Miao}}, \bibinfo {author}
  {\bibfnamefont {Y.}~\bibnamefont {Wang}}, \ and\ \bibinfo {author}
  {\bibfnamefont {Y.}~\bibnamefont {Ma}},\ }\bibfield  {title} {\enquote
  {\bibinfo {title} {Atlas: A real-space finite-difference implementation of
  orbital-free density functional theory},}\ }\href@noop {} {\bibfield
  {journal} {\bibinfo  {journal} {Computer Physics Communications}\ }\textbf
  {\bibinfo {volume} {200}},\ \bibinfo {pages} {87 -- 95} (\bibinfo {year}
  {2016}{\natexlab{a}})}\BibitemShut {NoStop}%
\bibitem [{\citenamefont {Genova}(2018)}]{pbcpy}%
  \BibitemOpen
  \bibfield  {author} {\bibinfo {author} {\bibfnamefont {A.}~\bibnamefont
  {Genova}},\ }\href {https://pypi.org/project/pbcpy} {\enquote {\bibinfo
  {title} {Pbcpy: a python3 package providing some useful abstractions to deal
  with molecules and materials under periodic boundary conditions (pbc).}}\ }
  (\bibinfo {year} {2018})\BibitemShut {NoStop}%
\bibitem [{\citenamefont {Lehtola}\ \emph {et~al.}(2018)\citenamefont
  {Lehtola}, \citenamefont {Steigemann}, \citenamefont {Oliveira},\ and\
  \citenamefont {Marques}}]{Lehtola_2018}%
  \BibitemOpen
  \bibfield  {author} {\bibinfo {author} {\bibfnamefont {S.}~\bibnamefont
  {Lehtola}}, \bibinfo {author} {\bibfnamefont {C.}~\bibnamefont {Steigemann}},
  \bibinfo {author} {\bibfnamefont {M.~J.}\ \bibnamefont {Oliveira}}, \ and\
  \bibinfo {author} {\bibfnamefont {M.~A.}\ \bibnamefont {Marques}},\
  }\bibfield  {title} {\enquote {\bibinfo {title} {Recent developments in libxc
  {\textemdash} a comprehensive library of functionals for density functional
  theory},}\ }\href {\doibase 10.1016/j.softx.2017.11.002} {\bibfield
  {journal} {\bibinfo  {journal} {{SoftwareX}}\ }\textbf {\bibinfo {volume}
  {7}},\ \bibinfo {pages} {1--5} (\bibinfo {year} {2018})}\BibitemShut
  {NoStop}%
\bibitem [{\citenamefont {Banerjee}\ and\ \citenamefont
  {Harbola}(2000)}]{Banerjee_2000}%
  \BibitemOpen
  \bibfield  {author} {\bibinfo {author} {\bibfnamefont {A.}~\bibnamefont
  {Banerjee}}\ and\ \bibinfo {author} {\bibfnamefont {M.~K.}\ \bibnamefont
  {Harbola}},\ }\bibfield  {title} {\enquote {\bibinfo {title} {Hydrodynamic
  approach to time-dependent density functional theory: Response properties of
  metal clusters},}\ }\href {\doibase 10.1063/1.1290610} {\bibfield  {journal}
  {\bibinfo  {journal} {The Journal of Chemical Physics}\ }\textbf {\bibinfo
  {volume} {113}},\ \bibinfo {pages} {5614--5623} (\bibinfo {year}
  {2000})}\BibitemShut {NoStop}%
\bibitem [{\citenamefont {Tokatly}\ and\ \citenamefont
  {Pankratov}(1999)}]{Tokatly_1999}%
  \BibitemOpen
  \bibfield  {author} {\bibinfo {author} {\bibfnamefont {I.}~\bibnamefont
  {Tokatly}}\ and\ \bibinfo {author} {\bibfnamefont {O.}~\bibnamefont
  {Pankratov}},\ }\bibfield  {title} {\enquote {\bibinfo {title} {Hydrodynamic
  theory of an electron gas},}\ }\href {\doibase 10.1103/physrevb.60.15550}
  {\bibfield  {journal} {\bibinfo  {journal} {Physical Review B}\ }\textbf
  {\bibinfo {volume} {60}},\ \bibinfo {pages} {15550--15553} (\bibinfo {year}
  {1999})}\BibitemShut {NoStop}%
\bibitem [{\citenamefont {Banerjee}\ and\ \citenamefont
  {Harbola}(2008)}]{Banerjee_2008}%
  \BibitemOpen
  \bibfield  {author} {\bibinfo {author} {\bibfnamefont {A.}~\bibnamefont
  {Banerjee}}\ and\ \bibinfo {author} {\bibfnamefont {M.~K.}\ \bibnamefont
  {Harbola}},\ }\bibfield  {title} {\enquote {\bibinfo {title} {Hydrodynamical
  approach to collective oscillations in metal clusters},}\ }\href {\doibase
  10.1016/j.physleta.2007.12.046} {\bibfield  {journal} {\bibinfo  {journal}
  {Physics Letters A}\ }\textbf {\bibinfo {volume} {372}},\ \bibinfo {pages}
  {2881--2886} (\bibinfo {year} {2008})}\BibitemShut {NoStop}%
\bibitem [{\citenamefont {Zaremba}\ and\ \citenamefont
  {Tso}(1994)}]{Zaremba_1994}%
  \BibitemOpen
  \bibfield  {author} {\bibinfo {author} {\bibfnamefont {E.}~\bibnamefont
  {Zaremba}}\ and\ \bibinfo {author} {\bibfnamefont {H.~C.}\ \bibnamefont
  {Tso}},\ }\bibfield  {title} {\enquote {\bibinfo {title}
  {Thomas-fermi-dirac-von weizs{\"a}cker hydrodynamics in parabolic wells},}\
  }\href {\doibase 10.1103/physrevb.49.8147} {\bibfield  {journal} {\bibinfo
  {journal} {Physical Review B}\ }\textbf {\bibinfo {volume} {49}},\ \bibinfo
  {pages} {8147--8162} (\bibinfo {year} {1994})}\BibitemShut {NoStop}%
\bibitem [{\citenamefont {Larsen}\ \emph {et~al.}(2017)\citenamefont {Larsen},
  \citenamefont {Mortensen}, \citenamefont {Blomqvist}, \citenamefont
  {Castelli}, \citenamefont {Christensen}, \citenamefont {Du{\l}ak},
  \citenamefont {Friis}, \citenamefont {Groves}, \citenamefont {Hammer},
  \citenamefont {Hargus}, \citenamefont {Hermes}, \citenamefont {Jennings},
  \citenamefont {Jensen}, \citenamefont {Kermode}, \citenamefont {Kitchin},
  \citenamefont {Kolsbjerg}, \citenamefont {Kubal}, \citenamefont {Kaasbjerg},
  \citenamefont {Lysgaard}, \citenamefont {Maronsson}, \citenamefont {Maxson},
  \citenamefont {Olsen}, \citenamefont {Pastewka}, \citenamefont {Peterson},
  \citenamefont {Rostgaard}, \citenamefont {Schi{\o}tz}, \citenamefont
  {Schütt}, \citenamefont {Strange}, \citenamefont {Thygesen}, \citenamefont
  {Vegge}, \citenamefont {Vilhelmsen}, \citenamefont {Walter}, \citenamefont
  {Zeng},\ and\ \citenamefont {Jacobsen}}]{Hjorth_Larsen_2017}%
  \BibitemOpen
  \bibfield  {author} {\bibinfo {author} {\bibfnamefont {A.~H.}\ \bibnamefont
  {Larsen}}, \bibinfo {author} {\bibfnamefont {J.~J.}\ \bibnamefont
  {Mortensen}}, \bibinfo {author} {\bibfnamefont {J.}~\bibnamefont
  {Blomqvist}}, \bibinfo {author} {\bibfnamefont {I.~E.}\ \bibnamefont
  {Castelli}}, \bibinfo {author} {\bibfnamefont {R.}~\bibnamefont
  {Christensen}}, \bibinfo {author} {\bibfnamefont {M.}~\bibnamefont
  {Du{\l}ak}}, \bibinfo {author} {\bibfnamefont {J.}~\bibnamefont {Friis}},
  \bibinfo {author} {\bibfnamefont {M.~N.}\ \bibnamefont {Groves}}, \bibinfo
  {author} {\bibfnamefont {B.}~\bibnamefont {Hammer}}, \bibinfo {author}
  {\bibfnamefont {C.}~\bibnamefont {Hargus}}, \bibinfo {author} {\bibfnamefont
  {E.~D.}\ \bibnamefont {Hermes}}, \bibinfo {author} {\bibfnamefont {P.~C.}\
  \bibnamefont {Jennings}}, \bibinfo {author} {\bibfnamefont {P.~B.}\
  \bibnamefont {Jensen}}, \bibinfo {author} {\bibfnamefont {J.}~\bibnamefont
  {Kermode}}, \bibinfo {author} {\bibfnamefont {J.~R.}\ \bibnamefont
  {Kitchin}}, \bibinfo {author} {\bibfnamefont {E.~L.}\ \bibnamefont
  {Kolsbjerg}}, \bibinfo {author} {\bibfnamefont {J.}~\bibnamefont {Kubal}},
  \bibinfo {author} {\bibfnamefont {K.}~\bibnamefont {Kaasbjerg}}, \bibinfo
  {author} {\bibfnamefont {S.}~\bibnamefont {Lysgaard}}, \bibinfo {author}
  {\bibfnamefont {J.~B.}\ \bibnamefont {Maronsson}}, \bibinfo {author}
  {\bibfnamefont {T.}~\bibnamefont {Maxson}}, \bibinfo {author} {\bibfnamefont
  {T.}~\bibnamefont {Olsen}}, \bibinfo {author} {\bibfnamefont
  {L.}~\bibnamefont {Pastewka}}, \bibinfo {author} {\bibfnamefont
  {A.}~\bibnamefont {Peterson}}, \bibinfo {author} {\bibfnamefont
  {C.}~\bibnamefont {Rostgaard}}, \bibinfo {author} {\bibfnamefont
  {J.}~\bibnamefont {Schi{\o}tz}}, \bibinfo {author} {\bibfnamefont
  {O.}~\bibnamefont {Schütt}}, \bibinfo {author} {\bibfnamefont
  {M.}~\bibnamefont {Strange}}, \bibinfo {author} {\bibfnamefont {K.~S.}\
  \bibnamefont {Thygesen}}, \bibinfo {author} {\bibfnamefont {T.}~\bibnamefont
  {Vegge}}, \bibinfo {author} {\bibfnamefont {L.}~\bibnamefont {Vilhelmsen}},
  \bibinfo {author} {\bibfnamefont {M.}~\bibnamefont {Walter}}, \bibinfo
  {author} {\bibfnamefont {Z.}~\bibnamefont {Zeng}}, \ and\ \bibinfo {author}
  {\bibfnamefont {K.~W.}\ \bibnamefont {Jacobsen}},\ }\bibfield  {title}
  {\enquote {\bibinfo {title} {The atomic simulation environment{\textemdash}a
  python library for working with atoms},}\ }\href {\doibase
  10.1088/1361-648x/aa680e} {\bibfield  {journal} {\bibinfo  {journal} {Journal
  of Physics: Condensed Matter}\ }\textbf {\bibinfo {volume} {29}},\ \bibinfo
  {pages} {273002} (\bibinfo {year} {2017})}\BibitemShut {NoStop}%
\bibitem [{\citenamefont {Darden}, \citenamefont {York},\ and\ \citenamefont
  {Pedersen}(1993)}]{Darden_1993}%
  \BibitemOpen
  \bibfield  {author} {\bibinfo {author} {\bibfnamefont {T.}~\bibnamefont
  {Darden}}, \bibinfo {author} {\bibfnamefont {D.}~\bibnamefont {York}}, \ and\
  \bibinfo {author} {\bibfnamefont {L.}~\bibnamefont {Pedersen}},\ }\bibfield
  {title} {\enquote {\bibinfo {title} {Particle mesh ewald: An n$\cdot$log(n)
  method for ewald sums in large systems},}\ }\href
  {https://doi.org/10.1063%2F1.464397} {\bibfield  {journal} {\bibinfo
  {journal} {J. Comp. Phys.}\ }\textbf {\bibinfo {volume} {98}},\ \bibinfo
  {pages} {10089--10092} (\bibinfo {year} {1993})}\BibitemShut {NoStop}%
\bibitem [{\citenamefont {Essmann}\ \emph {et~al.}(1995)\citenamefont
  {Essmann}, \citenamefont {Perera}, \citenamefont {Berkowitz}, \citenamefont
  {Darden}, \citenamefont {Lee},\ and\ \citenamefont {Pedersen}}]{Essmann}%
  \BibitemOpen
  \bibfield  {author} {\bibinfo {author} {\bibfnamefont {U.}~\bibnamefont
  {Essmann}}, \bibinfo {author} {\bibfnamefont {L.}~\bibnamefont {Perera}},
  \bibinfo {author} {\bibfnamefont {M.~L.}\ \bibnamefont {Berkowitz}}, \bibinfo
  {author} {\bibfnamefont {T.}~\bibnamefont {Darden}}, \bibinfo {author}
  {\bibfnamefont {H.}~\bibnamefont {Lee}}, \ and\ \bibinfo {author}
  {\bibfnamefont {L.~G.}\ \bibnamefont {Pedersen}},\ }\bibfield  {title}
  {\enquote {\bibinfo {title} {A smooth particle mesh ewald method},}\ }\href
  {\doibase http://dx.doi.org/10.1063/1.470117} {\bibfield  {journal} {\bibinfo
   {journal} {The Journal of Chemical Physics}\ }\textbf {\bibinfo {volume}
  {103}},\ \bibinfo {pages} {8577--8593} (\bibinfo {year} {1995})}\BibitemShut
  {NoStop}%
\bibitem [{\citenamefont {Sun}\ \emph {et~al.}(2017)\citenamefont {Sun},
  \citenamefont {Berkelbach}, \citenamefont {Blunt}, \citenamefont {Booth},
  \citenamefont {Guo}, \citenamefont {Li}, \citenamefont {Liu}, \citenamefont
  {McClain}, \citenamefont {Sayfutyarova}, \citenamefont {Sharma},
  \citenamefont {Wouters},\ and\ \citenamefont {Chan}}]{PYSCF}%
  \BibitemOpen
  \bibfield  {author} {\bibinfo {author} {\bibfnamefont {Q.}~\bibnamefont
  {Sun}}, \bibinfo {author} {\bibfnamefont {T.~C.}\ \bibnamefont {Berkelbach}},
  \bibinfo {author} {\bibfnamefont {N.~S.}\ \bibnamefont {Blunt}}, \bibinfo
  {author} {\bibfnamefont {G.~H.}\ \bibnamefont {Booth}}, \bibinfo {author}
  {\bibfnamefont {S.}~\bibnamefont {Guo}}, \bibinfo {author} {\bibfnamefont
  {Z.}~\bibnamefont {Li}}, \bibinfo {author} {\bibfnamefont {J.}~\bibnamefont
  {Liu}}, \bibinfo {author} {\bibfnamefont {J.~D.}\ \bibnamefont {McClain}},
  \bibinfo {author} {\bibfnamefont {E.~R.}\ \bibnamefont {Sayfutyarova}},
  \bibinfo {author} {\bibfnamefont {S.}~\bibnamefont {Sharma}}, \bibinfo
  {author} {\bibfnamefont {S.}~\bibnamefont {Wouters}}, \ and\ \bibinfo
  {author} {\bibfnamefont {G.~K.}\ \bibnamefont {Chan}},\ }\href {\doibase
  10.1002/wcms.1340} {\enquote {\bibinfo {title} {Pyscf: the python‐based
  simulations of chemistry framework},}\ } (\bibinfo {year} {2017})\BibitemShut
  {NoStop}%
\bibitem [{\citenamefont {Smith}\ \emph {et~al.}(2018)\citenamefont {Smith},
  \citenamefont {Burns}, \citenamefont {Sirianni}, \citenamefont {Nascimento},
  \citenamefont {Kumar}, \citenamefont {James}, \citenamefont {Schriber},
  \citenamefont {Zhang}, \citenamefont {Zhang}, \citenamefont {Abbott},
  \citenamefont {Berquist}, \citenamefont {Lechner}, \citenamefont {Cunha},
  \citenamefont {Heide}, \citenamefont {Waldrop}, \citenamefont {Takeshita},
  \citenamefont {Alenaizan}, \citenamefont {Neuhauser}, \citenamefont {King},
  \citenamefont {Simmonett}, \citenamefont {Turney}, \citenamefont {Schaefer},
  \citenamefont {Evangelista}, \citenamefont {DePrince}, \citenamefont
  {Crawford}, \citenamefont {Patkowski},\ and\ \citenamefont
  {Sherrill}}]{Smith_2018}%
  \BibitemOpen
  \bibfield  {author} {\bibinfo {author} {\bibfnamefont {D.~G.~A.}\
  \bibnamefont {Smith}}, \bibinfo {author} {\bibfnamefont {L.~A.}\ \bibnamefont
  {Burns}}, \bibinfo {author} {\bibfnamefont {D.~A.}\ \bibnamefont {Sirianni}},
  \bibinfo {author} {\bibfnamefont {D.~R.}\ \bibnamefont {Nascimento}},
  \bibinfo {author} {\bibfnamefont {A.}~\bibnamefont {Kumar}}, \bibinfo
  {author} {\bibfnamefont {A.~M.}\ \bibnamefont {James}}, \bibinfo {author}
  {\bibfnamefont {J.~B.}\ \bibnamefont {Schriber}}, \bibinfo {author}
  {\bibfnamefont {T.}~\bibnamefont {Zhang}}, \bibinfo {author} {\bibfnamefont
  {B.}~\bibnamefont {Zhang}}, \bibinfo {author} {\bibfnamefont {A.~S.}\
  \bibnamefont {Abbott}}, \bibinfo {author} {\bibfnamefont {E.~J.}\
  \bibnamefont {Berquist}}, \bibinfo {author} {\bibfnamefont {M.~H.}\
  \bibnamefont {Lechner}}, \bibinfo {author} {\bibfnamefont {L.~A.}\
  \bibnamefont {Cunha}}, \bibinfo {author} {\bibfnamefont {A.~G.}\ \bibnamefont
  {Heide}}, \bibinfo {author} {\bibfnamefont {J.~M.}\ \bibnamefont {Waldrop}},
  \bibinfo {author} {\bibfnamefont {T.~Y.}\ \bibnamefont {Takeshita}}, \bibinfo
  {author} {\bibfnamefont {A.}~\bibnamefont {Alenaizan}}, \bibinfo {author}
  {\bibfnamefont {D.}~\bibnamefont {Neuhauser}}, \bibinfo {author}
  {\bibfnamefont {R.~A.}\ \bibnamefont {King}}, \bibinfo {author}
  {\bibfnamefont {A.~C.}\ \bibnamefont {Simmonett}}, \bibinfo {author}
  {\bibfnamefont {J.~M.}\ \bibnamefont {Turney}}, \bibinfo {author}
  {\bibfnamefont {H.~F.}\ \bibnamefont {Schaefer}}, \bibinfo {author}
  {\bibfnamefont {F.~A.}\ \bibnamefont {Evangelista}}, \bibinfo {author}
  {\bibfnamefont {A.~E.}\ \bibnamefont {DePrince}}, \bibinfo {author}
  {\bibfnamefont {T.~D.}\ \bibnamefont {Crawford}}, \bibinfo {author}
  {\bibfnamefont {K.}~\bibnamefont {Patkowski}}, \ and\ \bibinfo {author}
  {\bibfnamefont {C.~D.}\ \bibnamefont {Sherrill}},\ }\bibfield  {title}
  {\enquote {\bibinfo {title} {Psi4numpy: An interactive quantum chemistry
  programming environment for reference implementations and rapid
  development},}\ }\href {\doibase 10.1021/acs.jctc.8b00286} {\bibfield
  {journal} {\bibinfo  {journal} {Journal of Chemical Theory and Computation}\
  }\textbf {\bibinfo {volume} {14}},\ \bibinfo {pages} {3504--3511} (\bibinfo
  {year} {2018})}\BibitemShut {NoStop}%
\bibitem [{dft(2020{\natexlab{a}})}]{dftpyweb}%
  \BibitemOpen
  \href {https://dftpy.rutgers.edu} {\enquote {\bibinfo {title} {Dftpy manual,
  available at: https://dftpy.rutgers.edu},}\ } (\bibinfo {year}
  {2020}{\natexlab{a}})\BibitemShut {NoStop}%
\bibitem [{dft(2020{\natexlab{b}})}]{dftpygit}%
  \BibitemOpen
  \href {https://gitlab.com/pavanello-research-group/dftpy} {\enquote {\bibinfo
  {title} {Dftpy, available at:
  https://gitlab.com/pavanello-research-group/dftpy},}\ } (\bibinfo {year}
  {2020}{\natexlab{b}})\BibitemShut {NoStop}%
\bibitem [{\citenamefont {Huang}\ and\ \citenamefont
  {Carter}(2010)}]{PhysRevB.81.045206}%
  \BibitemOpen
  \bibfield  {author} {\bibinfo {author} {\bibfnamefont {C.}~\bibnamefont
  {Huang}}\ and\ \bibinfo {author} {\bibfnamefont {E.~A.}\ \bibnamefont
  {Carter}},\ }\bibfield  {title} {\enquote {\bibinfo {title} {Nonlocal
  orbital-free kinetic energy density functional for semiconductors},}\ }\href
  {\doibase 10.1103/PhysRevB.81.045206} {\bibfield  {journal} {\bibinfo
  {journal} {Phys. Rev. B}\ }\textbf {\bibinfo {volume} {81}},\ \bibinfo
  {pages} {045206} (\bibinfo {year} {2010})}\BibitemShut {NoStop}%
\bibitem [{\citenamefont {Constantin}, \citenamefont {Fabiano},\ and\
  \citenamefont {Sala}(2018)}]{Constantin_2018}%
  \BibitemOpen
  \bibfield  {author} {\bibinfo {author} {\bibfnamefont {L.~A.}\ \bibnamefont
  {Constantin}}, \bibinfo {author} {\bibfnamefont {E.}~\bibnamefont {Fabiano}},
  \ and\ \bibinfo {author} {\bibfnamefont {F.~D.}\ \bibnamefont {Sala}},\
  }\bibfield  {title} {\enquote {\bibinfo {title} {Nonlocal kinetic energy
  functional from the jellium-with-gap model: Applications to orbital-free
  density functional theory},}\ }\href
  {https://doi.org/10.1103%2Fphysrevb.97.205137} {\bibfield  {journal}
  {\bibinfo  {journal} {Physical Review B}\ }\textbf {\bibinfo {volume} {97}}
  (\bibinfo {year} {2018})}\BibitemShut {NoStop}%
\bibitem [{\citenamefont {Luo}, \citenamefont {Karasiev},\ and\ \citenamefont
  {Trickey}(2018)}]{luo2018simple}%
  \BibitemOpen
  \bibfield  {author} {\bibinfo {author} {\bibfnamefont {K.}~\bibnamefont
  {Luo}}, \bibinfo {author} {\bibfnamefont {V.~V.}\ \bibnamefont {Karasiev}}, \
  and\ \bibinfo {author} {\bibfnamefont {S.}~\bibnamefont {Trickey}},\
  }\bibfield  {title} {\enquote {\bibinfo {title} {A simple generalized
  gradient approximation for the noninteracting kinetic energy density
  functional},}\ }\href@noop {} {\bibfield  {journal} {\bibinfo  {journal}
  {Phys. Rev. B}\ }\textbf {\bibinfo {volume} {98}},\ \bibinfo {pages} {041111}
  (\bibinfo {year} {2018})}\BibitemShut {NoStop}%
\bibitem [{\citenamefont {Constantin}\ \emph {et~al.}(2017)\citenamefont
  {Constantin}, \citenamefont {Fabiano}, \citenamefont {{\'S}miga},\ and\
  \citenamefont {Della~Sala}}]{constantin2017}%
  \BibitemOpen
  \bibfield  {author} {\bibinfo {author} {\bibfnamefont {L.~A.}\ \bibnamefont
  {Constantin}}, \bibinfo {author} {\bibfnamefont {E.}~\bibnamefont {Fabiano}},
  \bibinfo {author} {\bibfnamefont {S.}~\bibnamefont {{\'S}miga}}, \ and\
  \bibinfo {author} {\bibfnamefont {F.}~\bibnamefont {Della~Sala}},\ }\bibfield
   {title} {\enquote {\bibinfo {title} {Jellium-with-gap model applied to
  semilocal kinetic functionals},}\ }\href@noop {} {\bibfield  {journal}
  {\bibinfo  {journal} {Physical Review B}\ }\textbf {\bibinfo {volume} {95}},\
  \bibinfo {pages} {115153} (\bibinfo {year} {2017})}\BibitemShut {NoStop}%
\bibitem [{\citenamefont {Xu}\ \emph {et~al.}(2020)\citenamefont {Xu},
  \citenamefont {Lv}, \citenamefont {Wang},\ and\ \citenamefont
  {Ma}}]{xu2020nonlocal}%
  \BibitemOpen
  \bibfield  {author} {\bibinfo {author} {\bibfnamefont {Q.}~\bibnamefont
  {Xu}}, \bibinfo {author} {\bibfnamefont {J.}~\bibnamefont {Lv}}, \bibinfo
  {author} {\bibfnamefont {Y.}~\bibnamefont {Wang}}, \ and\ \bibinfo {author}
  {\bibfnamefont {Y.}~\bibnamefont {Ma}},\ }\bibfield  {title} {\enquote
  {\bibinfo {title} {Nonlocal kinetic energy density functionals for isolated
  systems obtained via local density approximation kernels},}\ }\href@noop {}
  {\bibfield  {journal} {\bibinfo  {journal} {Physical Review B}\ }\textbf
  {\bibinfo {volume} {101}},\ \bibinfo {pages} {045110} (\bibinfo {year}
  {2020})}\BibitemShut {NoStop}%
\bibitem [{\citenamefont {Mi}, \citenamefont {Genova},\ and\ \citenamefont
  {Pavanello}(2018)}]{mi2018nonlocal}%
  \BibitemOpen
  \bibfield  {author} {\bibinfo {author} {\bibfnamefont {W.}~\bibnamefont
  {Mi}}, \bibinfo {author} {\bibfnamefont {A.}~\bibnamefont {Genova}}, \ and\
  \bibinfo {author} {\bibfnamefont {M.}~\bibnamefont {Pavanello}},\ }\bibfield
  {title} {\enquote {\bibinfo {title} {Nonlocal kinetic energy functionals by
  functional integration},}\ }\href@noop {} {\bibfield  {journal} {\bibinfo
  {journal} {J. Comp. Phys.}\ }\textbf {\bibinfo {volume} {148}},\ \bibinfo
  {pages} {184107} (\bibinfo {year} {2018})}\BibitemShut {NoStop}%
\bibitem [{\citenamefont {Wang}\ and\ \citenamefont
  {Teter}(1992)}]{PhysRevB.45.13196}%
  \BibitemOpen
  \bibfield  {author} {\bibinfo {author} {\bibfnamefont {L.-W.}\ \bibnamefont
  {Wang}}\ and\ \bibinfo {author} {\bibfnamefont {M.~P.}\ \bibnamefont
  {Teter}},\ }\bibfield  {title} {\enquote {\bibinfo {title} {Kinetic-energy
  functional of the electron density},}\ }\href {\doibase
  10.1103/PhysRevB.45.13196} {\bibfield  {journal} {\bibinfo  {journal} {Phys.
  Rev. B}\ }\textbf {\bibinfo {volume} {45}},\ \bibinfo {pages} {13196--13220}
  (\bibinfo {year} {1992})}\BibitemShut {NoStop}%
\bibitem [{\citenamefont {Wang}, \citenamefont {Govind},\ and\ \citenamefont
  {Carter}(1999)}]{PhysRevB.60.16350}%
  \BibitemOpen
  \bibfield  {author} {\bibinfo {author} {\bibfnamefont {Y.~A.}\ \bibnamefont
  {Wang}}, \bibinfo {author} {\bibfnamefont {N.}~\bibnamefont {Govind}}, \ and\
  \bibinfo {author} {\bibfnamefont {E.~A.}\ \bibnamefont {Carter}},\ }\bibfield
   {title} {\enquote {\bibinfo {title} {Orbital-free kinetic-energy density
  functionals with a density-dependent kernel},}\ }\href {\doibase
  10.1103/PhysRevB.60.16350} {\bibfield  {journal} {\bibinfo  {journal} {Phys.
  Rev. B}\ }\textbf {\bibinfo {volume} {60}},\ \bibinfo {pages} {16350--16358}
  (\bibinfo {year} {1999})}\BibitemShut {NoStop}%
\bibitem [{\citenamefont {Hestenes}\ and\ \citenamefont
  {Stiefel}(1952)}]{hestenes1952methods}%
  \BibitemOpen
  \bibfield  {author} {\bibinfo {author} {\bibfnamefont {M.~R.}\ \bibnamefont
  {Hestenes}}\ and\ \bibinfo {author} {\bibfnamefont {E.}~\bibnamefont
  {Stiefel}},\ }\href@noop {} {\emph {\bibinfo {title} {Methods of conjugate
  gradients for solving linear systems}}},\ Vol.~\bibinfo {volume} {49}\
  (\bibinfo  {publisher} {NBS Washington, DC},\ \bibinfo {year}
  {1952})\BibitemShut {NoStop}%
\bibitem [{\citenamefont {Fletcher}\ and\ \citenamefont
  {Reeves}(1964)}]{fletcher1964function}%
  \BibitemOpen
  \bibfield  {author} {\bibinfo {author} {\bibfnamefont {R.}~\bibnamefont
  {Fletcher}}\ and\ \bibinfo {author} {\bibfnamefont {C.~M.}\ \bibnamefont
  {Reeves}},\ }\bibfield  {title} {\enquote {\bibinfo {title} {Function
  minimization by conjugate gradients},}\ }\href@noop {} {\bibfield  {journal}
  {\bibinfo  {journal} {The computer journal}\ }\textbf {\bibinfo {volume}
  {7}},\ \bibinfo {pages} {149--154} (\bibinfo {year} {1964})}\BibitemShut
  {NoStop}%
\bibitem [{\citenamefont {Polak}\ and\ \citenamefont
  {Ribiere}(1969)}]{polak1969note}%
  \BibitemOpen
  \bibfield  {author} {\bibinfo {author} {\bibfnamefont {E.}~\bibnamefont
  {Polak}}\ and\ \bibinfo {author} {\bibfnamefont {G.}~\bibnamefont
  {Ribiere}},\ }\bibfield  {title} {\enquote {\bibinfo {title} {Note sur la
  convergence de m{\'e}thodes de directions conjugu{\'e}es},}\ }\href@noop {}
  {\bibfield  {journal} {\bibinfo  {journal} {ESAIM: Mathematical Modelling and
  Numerical Analysis-Mod{\'e}lisation Math{\'e}matique et Analyse
  Num{\'e}rique}\ }\textbf {\bibinfo {volume} {3}},\ \bibinfo {pages} {35--43}
  (\bibinfo {year} {1969})}\BibitemShut {NoStop}%
\bibitem [{\citenamefont {Polyak}(1969)}]{polyak1969conjugate}%
  \BibitemOpen
  \bibfield  {author} {\bibinfo {author} {\bibfnamefont {B.~T.}\ \bibnamefont
  {Polyak}},\ }\bibfield  {title} {\enquote {\bibinfo {title} {The conjugate
  gradient method in extremal problems},}\ }\href@noop {} {\bibfield  {journal}
  {\bibinfo  {journal} {USSR Computational Mathematics and Mathematical
  Physics}\ }\textbf {\bibinfo {volume} {9}},\ \bibinfo {pages} {94--112}
  (\bibinfo {year} {1969})}\BibitemShut {NoStop}%
\bibitem [{\citenamefont {Fletcher}(1980)}]{fletcher1980practical}%
  \BibitemOpen
  \bibfield  {author} {\bibinfo {author} {\bibfnamefont {R.}~\bibnamefont
  {Fletcher}},\ }\href@noop {} {\emph {\bibinfo {title} {Practical Methods Of
  Optimization: Vol. 1 Unconstrained Optimization}}}\ (\bibinfo  {publisher}
  {John Wiley \& Sons},\ \bibinfo {year} {1980})\BibitemShut {NoStop}%
\bibitem [{\citenamefont {Liu}\ and\ \citenamefont
  {Storey}(1991)}]{liu1991efficient}%
  \BibitemOpen
  \bibfield  {author} {\bibinfo {author} {\bibfnamefont {Y.}~\bibnamefont
  {Liu}}\ and\ \bibinfo {author} {\bibfnamefont {C.}~\bibnamefont {Storey}},\
  }\bibfield  {title} {\enquote {\bibinfo {title} {Efficient generalized
  conjugate gradient algorithms, part 1: theory},}\ }\href@noop {} {\bibfield
  {journal} {\bibinfo  {journal} {Journal of optimization theory and
  applications}\ }\textbf {\bibinfo {volume} {69}},\ \bibinfo {pages}
  {129--137} (\bibinfo {year} {1991})}\BibitemShut {NoStop}%
\bibitem [{\citenamefont {Dai}\ and\ \citenamefont
  {Yuan}(1999)}]{dai1999nonlinear}%
  \BibitemOpen
  \bibfield  {author} {\bibinfo {author} {\bibfnamefont {Y.-H.}\ \bibnamefont
  {Dai}}\ and\ \bibinfo {author} {\bibfnamefont {Y.}~\bibnamefont {Yuan}},\
  }\bibfield  {title} {\enquote {\bibinfo {title} {A nonlinear conjugate
  gradient method with a strong global convergence property},}\ }\href@noop {}
  {\bibfield  {journal} {\bibinfo  {journal} {SIAM Journal on optimization}\
  }\textbf {\bibinfo {volume} {10}},\ \bibinfo {pages} {177--182} (\bibinfo
  {year} {1999})}\BibitemShut {NoStop}%
\bibitem [{\citenamefont {Liu}\ and\ \citenamefont
  {Nocedal}(1989)}]{liu1989limited}%
  \BibitemOpen
  \bibfield  {author} {\bibinfo {author} {\bibfnamefont {D.~C.}\ \bibnamefont
  {Liu}}\ and\ \bibinfo {author} {\bibfnamefont {J.}~\bibnamefont {Nocedal}},\
  }\bibfield  {title} {\enquote {\bibinfo {title} {On the limited memory bfgs
  method for large scale optimization},}\ }\href@noop {} {\bibfield  {journal}
  {\bibinfo  {journal} {Mathematical programming}\ }\textbf {\bibinfo {volume}
  {45}},\ \bibinfo {pages} {503--528} (\bibinfo {year} {1989})}\BibitemShut
  {NoStop}%
\bibitem [{\citenamefont {Nocedal}\ and\ \citenamefont
  {Wright}(2006)}]{nocedal2006numerical}%
  \BibitemOpen
  \bibfield  {author} {\bibinfo {author} {\bibfnamefont {J.}~\bibnamefont
  {Nocedal}}\ and\ \bibinfo {author} {\bibfnamefont {S.}~\bibnamefont
  {Wright}},\ }\href@noop {} {\emph {\bibinfo {title} {Numerical
  optimization}}}\ (\bibinfo  {publisher} {Springer Science \& Business
  Media},\ \bibinfo {year} {2006})\BibitemShut {NoStop}%
\bibitem [{\citenamefont {Jiang}\ and\ \citenamefont
  {Yang}(2004)}]{jiang2004conjugate}%
  \BibitemOpen
  \bibfield  {author} {\bibinfo {author} {\bibfnamefont {H.}~\bibnamefont
  {Jiang}}\ and\ \bibinfo {author} {\bibfnamefont {W.}~\bibnamefont {Yang}},\
  }\bibfield  {title} {\enquote {\bibinfo {title} {Conjugate-gradient
  optimization method for orbital-free density functional calculations},}\
  }\href@noop {} {\bibfield  {journal} {\bibinfo  {journal} {The Journal of
  chemical physics}\ }\textbf {\bibinfo {volume} {121}},\ \bibinfo {pages}
  {2030--2036} (\bibinfo {year} {2004})}\BibitemShut {NoStop}%
\bibitem [{\citenamefont {Frigo}\ and\ \citenamefont {Johnson}(2005)}]{FFTW05}%
  \BibitemOpen
  \bibfield  {author} {\bibinfo {author} {\bibfnamefont {M.}~\bibnamefont
  {Frigo}}\ and\ \bibinfo {author} {\bibfnamefont {S.~G.}\ \bibnamefont
  {Johnson}},\ }\bibfield  {title} {\enquote {\bibinfo {title} {The design and
  implementation of {FFTW3}},}\ }\href@noop {} {\bibfield  {journal} {\bibinfo
  {journal} {Proceedings of the IEEE}\ }\textbf {\bibinfo {volume} {93}},\
  \bibinfo {pages} {216--231} (\bibinfo {year} {2005})},\ \bibinfo {note}
  {special issue on ``Program Generation, Optimization, and Platform
  Adaptation''}\BibitemShut {NoStop}%
\bibitem [{\citenamefont {Choly}\ and\ \citenamefont
  {Kaxiras}(2003)}]{choly2003fast}%
  \BibitemOpen
  \bibfield  {author} {\bibinfo {author} {\bibfnamefont {N.}~\bibnamefont
  {Choly}}\ and\ \bibinfo {author} {\bibfnamefont {E.}~\bibnamefont
  {Kaxiras}},\ }\bibfield  {title} {\enquote {\bibinfo {title} {Fast method for
  force computations in electronic structure calculations},}\ }\href@noop {}
  {\bibfield  {journal} {\bibinfo  {journal} {Physical Review B}\ }\textbf
  {\bibinfo {volume} {67}},\ \bibinfo {pages} {155101} (\bibinfo {year}
  {2003})}\BibitemShut {NoStop}%
\bibitem [{\citenamefont {Hung}\ and\ \citenamefont
  {Carter}(2009{\natexlab{a}})}]{hung2009accurate}%
  \BibitemOpen
  \bibfield  {author} {\bibinfo {author} {\bibfnamefont {L.}~\bibnamefont
  {Hung}}\ and\ \bibinfo {author} {\bibfnamefont {E.~A.}\ \bibnamefont
  {Carter}},\ }\bibfield  {title} {\enquote {\bibinfo {title} {Accurate
  simulations of metals at the mesoscale: Explicit treatment of 1 million atoms
  with quantum mechanics},}\ }\href@noop {} {\bibfield  {journal} {\bibinfo
  {journal} {Chemical Physics Letters}\ }\textbf {\bibinfo {volume} {475}},\
  \bibinfo {pages} {163--170} (\bibinfo {year}
  {2009}{\natexlab{a}})}\BibitemShut {NoStop}%
\bibitem [{\citenamefont {Ho}, \citenamefont {Lign{\`e}res},\ and\
  \citenamefont {Carter}(2008)}]{PROFESS1.0}%
  \BibitemOpen
  \bibfield  {author} {\bibinfo {author} {\bibfnamefont {G.~S.}\ \bibnamefont
  {Ho}}, \bibinfo {author} {\bibfnamefont {V.~L.}\ \bibnamefont
  {Lign{\`e}res}}, \ and\ \bibinfo {author} {\bibfnamefont {E.~A.}\
  \bibnamefont {Carter}},\ }\bibfield  {title} {\enquote {\bibinfo {title}
  {Introducing profess: A new program for orbital-free density functional
  theory calculations},}\ }\href {\doibase
  http://dx.doi.org/10.1016/j.cpc.2008.07.002} {\bibfield  {journal} {\bibinfo
  {journal} {Computer Physics Communications}\ }\textbf {\bibinfo {volume}
  {179}},\ \bibinfo {pages} {839 -- 854} (\bibinfo {year} {2008})}\BibitemShut
  {NoStop}%
\bibitem [{\citenamefont {Shao}\ \emph {et~al.}(2016)\citenamefont {Shao},
  \citenamefont {Mi}, \citenamefont {Xu}, \citenamefont {Wang},\ and\
  \citenamefont {Ma}}]{shao2016n}%
  \BibitemOpen
  \bibfield  {author} {\bibinfo {author} {\bibfnamefont {X.}~\bibnamefont
  {Shao}}, \bibinfo {author} {\bibfnamefont {W.}~\bibnamefont {Mi}}, \bibinfo
  {author} {\bibfnamefont {Q.}~\bibnamefont {Xu}}, \bibinfo {author}
  {\bibfnamefont {Y.}~\bibnamefont {Wang}}, \ and\ \bibinfo {author}
  {\bibfnamefont {Y.}~\bibnamefont {Ma}},\ }\bibfield  {title} {\enquote
  {\bibinfo {title} {O (n log n) scaling method to evaluate the ion--electron
  potential of crystalline solids},}\ }\href@noop {} {\bibfield  {journal}
  {\bibinfo  {journal} {J. Comp. Phys.}\ }\textbf {\bibinfo {volume} {145}},\
  \bibinfo {pages} {184110} (\bibinfo {year} {2016})}\BibitemShut {NoStop}%
\bibitem [{\citenamefont {Gomersall}(2016)}]{pyfftw}%
  \BibitemOpen
  \bibfield  {author} {\bibinfo {author} {\bibfnamefont {H.}~\bibnamefont
  {Gomersall}},\ }\href {https://doi.org/10.5281/zenodo.59508} {\enquote
  {\bibinfo {title} {pyfftw},}\ } (\bibinfo {year} {2016})\BibitemShut
  {NoStop}%
\bibitem [{\citenamefont {Opanchuk}(2019)}]{reikna}%
  \BibitemOpen
  \bibfield  {author} {\bibinfo {author} {\bibfnamefont {B.}~\bibnamefont
  {Opanchuk}},\ }\href {http://reikna.publicfields.net} {\enquote {\bibinfo
  {title} {Reikna, a pure python gpgpu library},}\ } (\bibinfo {year}
  {2019})\BibitemShut {NoStop}%
\bibitem [{\citenamefont {Abadi}\ \emph {et~al.}(2015)\citenamefont {Abadi},
  \citenamefont {Agarwal}, \citenamefont {Barham}, \citenamefont {Brevdo},
  \citenamefont {Chen}, \citenamefont {Citro}, \citenamefont {Corrado},
  \citenamefont {Davis}, \citenamefont {Dean}, \citenamefont {Devin},
  \citenamefont {Ghemawat}, \citenamefont {Goodfellow}, \citenamefont {Harp},
  \citenamefont {Irving}, \citenamefont {Isard}, \citenamefont {Jia},
  \citenamefont {Jozefowicz}, \citenamefont {Kaiser}, \citenamefont {Kudlur},
  \citenamefont {Levenberg}, \citenamefont {Man\'{e}}, \citenamefont {Monga},
  \citenamefont {Moore}, \citenamefont {Murray}, \citenamefont {Olah},
  \citenamefont {Schuster}, \citenamefont {Shlens}, \citenamefont {Steiner},
  \citenamefont {Sutskever}, \citenamefont {Talwar}, \citenamefont {Tucker},
  \citenamefont {Vanhoucke}, \citenamefont {Vasudevan}, \citenamefont
  {Vi\'{e}gas}, \citenamefont {Vinyals}, \citenamefont {Warden}, \citenamefont
  {Wattenberg}, \citenamefont {Wicke}, \citenamefont {Yu},\ and\ \citenamefont
  {Zheng}}]{tensorflow2015-whitepaper}%
  \BibitemOpen
  \bibfield  {author} {\bibinfo {author} {\bibfnamefont {M.}~\bibnamefont
  {Abadi}}, \bibinfo {author} {\bibfnamefont {A.}~\bibnamefont {Agarwal}},
  \bibinfo {author} {\bibfnamefont {P.}~\bibnamefont {Barham}}, \bibinfo
  {author} {\bibfnamefont {E.}~\bibnamefont {Brevdo}}, \bibinfo {author}
  {\bibfnamefont {Z.}~\bibnamefont {Chen}}, \bibinfo {author} {\bibfnamefont
  {C.}~\bibnamefont {Citro}}, \bibinfo {author} {\bibfnamefont {G.~S.}\
  \bibnamefont {Corrado}}, \bibinfo {author} {\bibfnamefont {A.}~\bibnamefont
  {Davis}}, \bibinfo {author} {\bibfnamefont {J.}~\bibnamefont {Dean}},
  \bibinfo {author} {\bibfnamefont {M.}~\bibnamefont {Devin}}, \bibinfo
  {author} {\bibfnamefont {S.}~\bibnamefont {Ghemawat}}, \bibinfo {author}
  {\bibfnamefont {I.}~\bibnamefont {Goodfellow}}, \bibinfo {author}
  {\bibfnamefont {A.}~\bibnamefont {Harp}}, \bibinfo {author} {\bibfnamefont
  {G.}~\bibnamefont {Irving}}, \bibinfo {author} {\bibfnamefont
  {M.}~\bibnamefont {Isard}}, \bibinfo {author} {\bibfnamefont
  {Y.}~\bibnamefont {Jia}}, \bibinfo {author} {\bibfnamefont {R.}~\bibnamefont
  {Jozefowicz}}, \bibinfo {author} {\bibfnamefont {L.}~\bibnamefont {Kaiser}},
  \bibinfo {author} {\bibfnamefont {M.}~\bibnamefont {Kudlur}}, \bibinfo
  {author} {\bibfnamefont {J.}~\bibnamefont {Levenberg}}, \bibinfo {author}
  {\bibfnamefont {D.}~\bibnamefont {Man\'{e}}}, \bibinfo {author}
  {\bibfnamefont {R.}~\bibnamefont {Monga}}, \bibinfo {author} {\bibfnamefont
  {S.}~\bibnamefont {Moore}}, \bibinfo {author} {\bibfnamefont
  {D.}~\bibnamefont {Murray}}, \bibinfo {author} {\bibfnamefont
  {C.}~\bibnamefont {Olah}}, \bibinfo {author} {\bibfnamefont {M.}~\bibnamefont
  {Schuster}}, \bibinfo {author} {\bibfnamefont {J.}~\bibnamefont {Shlens}},
  \bibinfo {author} {\bibfnamefont {B.}~\bibnamefont {Steiner}}, \bibinfo
  {author} {\bibfnamefont {I.}~\bibnamefont {Sutskever}}, \bibinfo {author}
  {\bibfnamefont {K.}~\bibnamefont {Talwar}}, \bibinfo {author} {\bibfnamefont
  {P.}~\bibnamefont {Tucker}}, \bibinfo {author} {\bibfnamefont
  {V.}~\bibnamefont {Vanhoucke}}, \bibinfo {author} {\bibfnamefont
  {V.}~\bibnamefont {Vasudevan}}, \bibinfo {author} {\bibfnamefont
  {F.}~\bibnamefont {Vi\'{e}gas}}, \bibinfo {author} {\bibfnamefont
  {O.}~\bibnamefont {Vinyals}}, \bibinfo {author} {\bibfnamefont
  {P.}~\bibnamefont {Warden}}, \bibinfo {author} {\bibfnamefont
  {M.}~\bibnamefont {Wattenberg}}, \bibinfo {author} {\bibfnamefont
  {M.}~\bibnamefont {Wicke}}, \bibinfo {author} {\bibfnamefont
  {Y.}~\bibnamefont {Yu}}, \ and\ \bibinfo {author} {\bibfnamefont
  {X.}~\bibnamefont {Zheng}},\ }\href {https://www.tensorflow.org/} {\enquote
  {\bibinfo {title} {{TensorFlow}: Large-scale machine learning on
  heterogeneous systems},}\ } (\bibinfo {year} {2015})\BibitemShut {NoStop}%
\bibitem [{\citenamefont {Huang}\ and\ \citenamefont
  {Carter}(2008)}]{huang2008}%
  \BibitemOpen
  \bibfield  {author} {\bibinfo {author} {\bibfnamefont {C.}~\bibnamefont
  {Huang}}\ and\ \bibinfo {author} {\bibfnamefont {E.~A.}\ \bibnamefont
  {Carter}},\ }\bibfield  {title} {\enquote {\bibinfo {title} {Transferable
  local pseudopotentials for magnesium, aluminum and silicon},}\ }\href@noop {}
  {\bibfield  {journal} {\bibinfo  {journal} {Physical Chemistry Chemical
  Physics}\ }\textbf {\bibinfo {volume} {10}},\ \bibinfo {pages} {7109--7120}
  (\bibinfo {year} {2008})}\BibitemShut {NoStop}%
\bibitem [{\citenamefont {Mi}\ \emph {et~al.}(2016{\natexlab{b}})\citenamefont
  {Mi}, \citenamefont {Zhang}, \citenamefont {Wang}, \citenamefont {Ma},\ and\
  \citenamefont {Miao}}]{mi2016first}%
  \BibitemOpen
  \bibfield  {author} {\bibinfo {author} {\bibfnamefont {W.}~\bibnamefont
  {Mi}}, \bibinfo {author} {\bibfnamefont {S.}~\bibnamefont {Zhang}}, \bibinfo
  {author} {\bibfnamefont {Y.}~\bibnamefont {Wang}}, \bibinfo {author}
  {\bibfnamefont {Y.}~\bibnamefont {Ma}}, \ and\ \bibinfo {author}
  {\bibfnamefont {M.}~\bibnamefont {Miao}},\ }\bibfield  {title} {\enquote
  {\bibinfo {title} {First-principle optimal local pseudopotentials
  construction via optimized effective potential method},}\ }\href@noop {}
  {\bibfield  {journal} {\bibinfo  {journal} {The Journal of chemical physics}\
  }\textbf {\bibinfo {volume} {144}},\ \bibinfo {pages} {134108} (\bibinfo
  {year} {2016}{\natexlab{b}})}\BibitemShut {NoStop}%
\bibitem [{\citenamefont {Perdew}\ and\ \citenamefont
  {Zunger}(1981)}]{Perdew1981}%
  \BibitemOpen
  \bibfield  {author} {\bibinfo {author} {\bibfnamefont {J.~P.}\ \bibnamefont
  {Perdew}}\ and\ \bibinfo {author} {\bibfnamefont {A.}~\bibnamefont
  {Zunger}},\ }\bibfield  {title} {\enquote {\bibinfo {title} {Self-interaction
  correction to density-functional approximations for many-electron systems},}\
  }\href {\doibase 10.1103/PhysRevB.23.5048} {\bibfield  {journal} {\bibinfo
  {journal} {Phys. Rev. B}\ }\textbf {\bibinfo {volume} {23}},\ \bibinfo
  {pages} {5048--5079} (\bibinfo {year} {1981})}\BibitemShut {NoStop}%
\bibitem [{\citenamefont {Krishtal}\ \emph {et~al.}(2015)\citenamefont
  {Krishtal}, \citenamefont {Sinha}, \citenamefont {Genova},\ and\
  \citenamefont {Pavanello}}]{krishtal2015subsystem}%
  \BibitemOpen
  \bibfield  {author} {\bibinfo {author} {\bibfnamefont {A.}~\bibnamefont
  {Krishtal}}, \bibinfo {author} {\bibfnamefont {D.}~\bibnamefont {Sinha}},
  \bibinfo {author} {\bibfnamefont {A.}~\bibnamefont {Genova}}, \ and\ \bibinfo
  {author} {\bibfnamefont {M.}~\bibnamefont {Pavanello}},\ }\bibfield  {title}
  {\enquote {\bibinfo {title} {Subsystem density-functional theory as an
  effective tool for modeling ground and excited states, their dynamics and
  many-body interactions},}\ }\href@noop {} {\bibfield  {journal} {\bibinfo
  {journal} {Journal of Physics: Condensed Matter}\ }\textbf {\bibinfo {volume}
  {27}},\ \bibinfo {pages} {183202} (\bibinfo {year} {2015})}\BibitemShut
  {NoStop}%
\bibitem [{\citenamefont {Hung}\ and\ \citenamefont
  {Carter}(2009{\natexlab{b}})}]{Hung2009163}%
  \BibitemOpen
  \bibfield  {author} {\bibinfo {author} {\bibfnamefont {L.}~\bibnamefont
  {Hung}}\ and\ \bibinfo {author} {\bibfnamefont {E.~A.}\ \bibnamefont
  {Carter}},\ }\bibfield  {title} {\enquote {\bibinfo {title} {Accurate
  simulations of metals at the mesoscale: Explicit treatment of 1 million atoms
  with quantum mechanics},}\ }\href {\doibase
  http://dx.doi.org/10.1016/j.cplett.2009.04.059} {\bibfield  {journal}
  {\bibinfo  {journal} {Chemical Physics Letters}\ }\textbf {\bibinfo {volume}
  {475}},\ \bibinfo {pages} {163 -- 170} (\bibinfo {year}
  {2009}{\natexlab{b}})}\BibitemShut {NoStop}%
\bibitem [{\citenamefont {Guelton}\ \emph {et~al.}(2015)\citenamefont
  {Guelton}, \citenamefont {Brunet}, \citenamefont {Amini}, \citenamefont
  {Merlini}, \citenamefont {Corbillon},\ and\ \citenamefont
  {Raynaud}}]{guelton2015pythran}%
  \BibitemOpen
  \bibfield  {author} {\bibinfo {author} {\bibfnamefont {S.}~\bibnamefont
  {Guelton}}, \bibinfo {author} {\bibfnamefont {P.}~\bibnamefont {Brunet}},
  \bibinfo {author} {\bibfnamefont {M.}~\bibnamefont {Amini}}, \bibinfo
  {author} {\bibfnamefont {A.}~\bibnamefont {Merlini}}, \bibinfo {author}
  {\bibfnamefont {X.}~\bibnamefont {Corbillon}}, \ and\ \bibinfo {author}
  {\bibfnamefont {A.}~\bibnamefont {Raynaud}},\ }\bibfield  {title} {\enquote
  {\bibinfo {title} {Pythran: Enabling static optimization of scientific python
  programs},}\ }\href@noop {} {\bibfield  {journal} {\bibinfo  {journal}
  {Computational Science \& Discovery}\ }\textbf {\bibinfo {volume} {8}},\
  \bibinfo {pages} {014001} (\bibinfo {year} {2015})}\BibitemShut {NoStop}%
\bibitem [{\citenamefont {Waseda}(1980)}]{waseda1980structure}%
  \BibitemOpen
  \bibfield  {author} {\bibinfo {author} {\bibfnamefont {Y.}~\bibnamefont
  {Waseda}},\ }\bibfield  {title} {\enquote {\bibinfo {title} {The structure of
  non-crystalline materials},}\ }\href@noop {} {\bibfield  {journal} {\bibinfo
  {journal} {Liguids and Amorphous Solids}\ } (\bibinfo {year}
  {1980})}\BibitemShut {NoStop}%
\bibitem [{\citenamefont {Allen}\ and\ \citenamefont
  {Tildesley}(2017)}]{allen2017computer}%
  \BibitemOpen
  \bibfield  {author} {\bibinfo {author} {\bibfnamefont {M.~P.}\ \bibnamefont
  {Allen}}\ and\ \bibinfo {author} {\bibfnamefont {D.~J.}\ \bibnamefont
  {Tildesley}},\ }\href@noop {} {\emph {\bibinfo {title} {Computer simulation
  of liquids}}}\ (\bibinfo  {publisher} {Oxford university press},\ \bibinfo
  {year} {2017})\BibitemShut {NoStop}%
\bibitem [{\citenamefont {Cirac{\`{\i}}}\ and\ \citenamefont
  {Sala}(2016)}]{Cirac__2016}%
  \BibitemOpen
  \bibfield  {author} {\bibinfo {author} {\bibfnamefont {C.}~\bibnamefont
  {Cirac{\`{\i}}}}\ and\ \bibinfo {author} {\bibfnamefont {F.~D.}\ \bibnamefont
  {Sala}},\ }\bibfield  {title} {\enquote {\bibinfo {title} {Quantum
  hydrodynamic theory for plasmonics: Impact of the electron density tail},}\
  }\href {https://doi.org/10.1103%2Fphysrevb.93.205405} {\bibfield  {journal}
  {\bibinfo  {journal} {Physical Review B}\ }\textbf {\bibinfo {volume} {93}}
  (\bibinfo {year} {2016})}\BibitemShut {NoStop}%
\bibitem [{\citenamefont {White}\ \emph {et~al.}(2013)\citenamefont {White},
  \citenamefont {Richardson}, \citenamefont {Crowley}, \citenamefont
  {Pattison}, \citenamefont {Harris},\ and\ \citenamefont
  {Gregori}}]{PhysRevLett.111.175002}%
  \BibitemOpen
  \bibfield  {author} {\bibinfo {author} {\bibfnamefont {T.~G.}\ \bibnamefont
  {White}}, \bibinfo {author} {\bibfnamefont {S.}~\bibnamefont {Richardson}},
  \bibinfo {author} {\bibfnamefont {B.~J.~B.}\ \bibnamefont {Crowley}},
  \bibinfo {author} {\bibfnamefont {L.~K.}\ \bibnamefont {Pattison}}, \bibinfo
  {author} {\bibfnamefont {J.~W.~O.}\ \bibnamefont {Harris}}, \ and\ \bibinfo
  {author} {\bibfnamefont {G.}~\bibnamefont {Gregori}},\ }\bibfield  {title}
  {\enquote {\bibinfo {title} {Orbital-free density-functional theory
  simulations of the dynamic structure factor of warm dense aluminum},}\ }\href
  {\doibase 10.1103/PhysRevLett.111.175002} {\bibfield  {journal} {\bibinfo
  {journal} {Phys. Rev. Lett.}\ }\textbf {\bibinfo {volume} {111}},\ \bibinfo
  {pages} {175002} (\bibinfo {year} {2013})}\BibitemShut {NoStop}%
\bibitem [{\citenamefont {Bennett}(1970)}]{Bennett_1970}%
  \BibitemOpen
  \bibfield  {author} {\bibinfo {author} {\bibfnamefont {A.~J.}\ \bibnamefont
  {Bennett}},\ }\bibfield  {title} {\enquote {\bibinfo {title} {Influence of
  the electron charge distribution on surface-plasmon dispersion},}\ }\href
  {\doibase 10.1103/physrevb.1.203} {\bibfield  {journal} {\bibinfo  {journal}
  {Physical Review B}\ }\textbf {\bibinfo {volume} {1}},\ \bibinfo {pages}
  {203--207} (\bibinfo {year} {1970})}\BibitemShut {NoStop}%
\bibitem [{\citenamefont {Liebsch}(1997)}]{Liebsch_1997}%
  \BibitemOpen
  \bibfield  {author} {\bibinfo {author} {\bibfnamefont {A.}~\bibnamefont
  {Liebsch}},\ }\href {\doibase 10.1007/978-1-4757-5107-9} {\emph {\bibinfo
  {title} {Electronic Excitations at Metal Surfaces}}}\ (\bibinfo  {publisher}
  {Springer {US}},\ \bibinfo {year} {1997})\BibitemShut {NoStop}%
\bibitem [{\citenamefont {Harbola}(1998)}]{Harbola_1998}%
  \BibitemOpen
  \bibfield  {author} {\bibinfo {author} {\bibfnamefont {M.~K.}\ \bibnamefont
  {Harbola}},\ }\bibfield  {title} {\enquote {\bibinfo {title} {Differential
  virial theorem and quantum fluid dynamics},}\ }\href {\doibase
  10.1103/physreva.58.1779} {\bibfield  {journal} {\bibinfo  {journal}
  {Physical Review A}\ }\textbf {\bibinfo {volume} {58}},\ \bibinfo {pages}
  {1779--1782} (\bibinfo {year} {1998})}\BibitemShut {NoStop}%
\bibitem [{\citenamefont {Banerjee}\ and\ \citenamefont
  {Harbola}(2002)}]{Banerjee_2002}%
  \BibitemOpen
  \bibfield  {author} {\bibinfo {author} {\bibfnamefont {A.}~\bibnamefont
  {Banerjee}}\ and\ \bibinfo {author} {\bibfnamefont {M.~K.}\ \bibnamefont
  {Harbola}},\ }\bibfield  {title} {\enquote {\bibinfo {title} {Calculation of
  van der waals coefficients in hydrodynamic approach to time-dependent density
  functional theory},}\ }\href {\doibase 10.1063/1.1510730} {\bibfield
  {journal} {\bibinfo  {journal} {The Journal of Chemical Physics}\ }\textbf
  {\bibinfo {volume} {117}},\ \bibinfo {pages} {7845--7851} (\bibinfo {year}
  {2002})}\BibitemShut {NoStop}%
\bibitem [{\citenamefont {Castro}, \citenamefont {Marques},\ and\ \citenamefont
  {Rubio}(2004)}]{cast2004}%
  \BibitemOpen
  \bibfield  {author} {\bibinfo {author} {\bibfnamefont {A.}~\bibnamefont
  {Castro}}, \bibinfo {author} {\bibfnamefont {M.~A.~L.}\ \bibnamefont
  {Marques}}, \ and\ \bibinfo {author} {\bibfnamefont {A.}~\bibnamefont
  {Rubio}},\ }\bibfield  {title} {\enquote {\bibinfo {title} {{Propagators for
  the time-dependent Kohn-Sham equations}},}\ }\href {\doibase
  10.1063/1.1774980} {\bibfield  {journal} {\bibinfo  {journal} {J. Comp.
  Phys.}\ }\textbf {\bibinfo {volume} {121}},\ \bibinfo {pages} {3425--3433}
  (\bibinfo {year} {2004})}\BibitemShut {NoStop}%
\bibitem [{\citenamefont {{von Weizs{\"a}cker}}(1935)}]{weiz1935}%
  \BibitemOpen
  \bibfield  {author} {\bibinfo {author} {\bibfnamefont {C.~F.}\ \bibnamefont
  {{von Weizs{\"a}cker}}},\ }\bibfield  {title} {\enquote {\bibinfo {title}
  {{Zur Theorie der Kernmassen}},}\ }\href@noop {} {\bibfield  {journal}
  {\bibinfo  {journal} {Z. Physik}\ }\textbf {\bibinfo {volume} {96}},\
  \bibinfo {pages} {431--458} (\bibinfo {year} {1935})}\BibitemShut {NoStop}%
\bibitem [{\citenamefont {Chan}, \citenamefont {Cohen},\ and\ \citenamefont
  {Handy}(2001)}]{chan2001thomas}%
  \BibitemOpen
  \bibfield  {author} {\bibinfo {author} {\bibfnamefont {G.~K.-L.}\
  \bibnamefont {Chan}}, \bibinfo {author} {\bibfnamefont {A.~J.}\ \bibnamefont
  {Cohen}}, \ and\ \bibinfo {author} {\bibfnamefont {N.~C.}\ \bibnamefont
  {Handy}},\ }\bibfield  {title} {\enquote {\bibinfo {title}
  {Thomas--fermi--dirac--von weizs{\"a}cker models in finite systems},}\ }\href
  {\doibase http://dx.doi.org/10.1063/1.1321308} {\bibfield  {journal}
  {\bibinfo  {journal} {The Journal of Chemical Physics}\ }\textbf {\bibinfo
  {volume} {114}},\ \bibinfo {pages} {631--638} (\bibinfo {year}
  {2001})}\BibitemShut {NoStop}%
\bibitem [{\citenamefont {Genova}\ \emph {et~al.}(2017)\citenamefont {Genova},
  \citenamefont {Ceresoli}, \citenamefont {Krishtal}, \citenamefont
  {Andreussi}, \citenamefont {DiStasio~Jr},\ and\ \citenamefont
  {Pavanello}}]{genova2017eqe}%
  \BibitemOpen
  \bibfield  {author} {\bibinfo {author} {\bibfnamefont {A.}~\bibnamefont
  {Genova}}, \bibinfo {author} {\bibfnamefont {D.}~\bibnamefont {Ceresoli}},
  \bibinfo {author} {\bibfnamefont {A.}~\bibnamefont {Krishtal}}, \bibinfo
  {author} {\bibfnamefont {O.}~\bibnamefont {Andreussi}}, \bibinfo {author}
  {\bibfnamefont {R.~A.}\ \bibnamefont {DiStasio~Jr}}, \ and\ \bibinfo {author}
  {\bibfnamefont {M.}~\bibnamefont {Pavanello}},\ }\bibfield  {title} {\enquote
  {\bibinfo {title} {eqe: An open-source density functional embedding theory
  code for the condensed phase},}\ }\href@noop {} {\bibfield  {journal}
  {\bibinfo  {journal} {International Journal of Quantum Chemistry}\ }\textbf
  {\bibinfo {volume} {117}},\ \bibinfo {pages} {e25401} (\bibinfo {year}
  {2017})}\BibitemShut {NoStop}%
\bibitem [{\citenamefont {Ong}\ \emph {et~al.}(2013)\citenamefont {Ong},
  \citenamefont {Richards}, \citenamefont {Jain}, \citenamefont {Hautier},
  \citenamefont {Kocher}, \citenamefont {Cholia}, \citenamefont {Gunter},
  \citenamefont {Chevrier}, \citenamefont {Persson},\ and\ \citenamefont
  {Ceder}}]{Ong_2013}%
  \BibitemOpen
  \bibfield  {author} {\bibinfo {author} {\bibfnamefont {S.~P.}\ \bibnamefont
  {Ong}}, \bibinfo {author} {\bibfnamefont {W.~D.}\ \bibnamefont {Richards}},
  \bibinfo {author} {\bibfnamefont {A.}~\bibnamefont {Jain}}, \bibinfo {author}
  {\bibfnamefont {G.}~\bibnamefont {Hautier}}, \bibinfo {author} {\bibfnamefont
  {M.}~\bibnamefont {Kocher}}, \bibinfo {author} {\bibfnamefont
  {S.}~\bibnamefont {Cholia}}, \bibinfo {author} {\bibfnamefont
  {D.}~\bibnamefont {Gunter}}, \bibinfo {author} {\bibfnamefont {V.~L.}\
  \bibnamefont {Chevrier}}, \bibinfo {author} {\bibfnamefont {K.~A.}\
  \bibnamefont {Persson}}, \ and\ \bibinfo {author} {\bibfnamefont
  {G.}~\bibnamefont {Ceder}},\ }\bibfield  {title} {\enquote {\bibinfo {title}
  {Python materials genomics (pymatgen): A robust, open-source python library
  for materials analysis},}\ }\href {\doibase 10.1016/j.commatsci.2012.10.028}
  {\bibfield  {journal} {\bibinfo  {journal} {Computational Materials Science}\
  }\textbf {\bibinfo {volume} {68}},\ \bibinfo {pages} {314--319} (\bibinfo
  {year} {2013})}\BibitemShut {NoStop}%
\bibitem [{\citenamefont {Jacob}\ \emph {et~al.}(2011)\citenamefont {Jacob},
  \citenamefont {Beyhan}, \citenamefont {Bulo}, \citenamefont {Gomes},
  \citenamefont {Götz}, \citenamefont {Kiewisch}, \citenamefont {Sikkema},\
  and\ \citenamefont {Visscher}}]{Jacob_2011}%
  \BibitemOpen
  \bibfield  {author} {\bibinfo {author} {\bibfnamefont {C.~R.}\ \bibnamefont
  {Jacob}}, \bibinfo {author} {\bibfnamefont {S.~M.}\ \bibnamefont {Beyhan}},
  \bibinfo {author} {\bibfnamefont {R.~E.}\ \bibnamefont {Bulo}}, \bibinfo
  {author} {\bibfnamefont {A.~S.~P.}\ \bibnamefont {Gomes}}, \bibinfo {author}
  {\bibfnamefont {A.~W.}\ \bibnamefont {Götz}}, \bibinfo {author}
  {\bibfnamefont {K.}~\bibnamefont {Kiewisch}}, \bibinfo {author}
  {\bibfnamefont {J.}~\bibnamefont {Sikkema}}, \ and\ \bibinfo {author}
  {\bibfnamefont {L.}~\bibnamefont {Visscher}},\ }\bibfield  {title} {\enquote
  {\bibinfo {title} {{PyADF} - a scripting framework for multiscale quantum
  chemistry},}\ }\href {\doibase 10.1002/jcc.21810} {\bibfield  {journal}
  {\bibinfo  {journal} {Journal of Computational Chemistry}\ }\textbf {\bibinfo
  {volume} {32}},\ \bibinfo {pages} {2328--2338} (\bibinfo {year}
  {2011})}\BibitemShut {NoStop}%
\bibitem [{\citenamefont {Muller}(2004)}]{pyquante}%
  \BibitemOpen
  \bibfield  {author} {\bibinfo {author} {\bibfnamefont {R.~P.}\ \bibnamefont
  {Muller}},\ }\href {http://pyquante.sourceforge.net/} {\enquote {\bibinfo
  {title} {{PyQuante: Python Quantum Chemistry.}}}\ } (\bibinfo {year}
  {2004})\BibitemShut {NoStop}%
\end{thebibliography}%
